\documentclass[acmsmall]{acmart}
\AtBeginDocument{%
  \providecommand\BibTeX{{%
    \normalfont B\kern-0.5em{\scshape i\kern-0.25em b}\kern-0.8em\TeX}}}

\setcopyright{acmlicensed}
\copyrightyear{2026}
\acmYear{2026}
\acmDOI{XXXXXXX.XXXXXXX}

\acmJournal{CSUR}
\acmVolume{x}
\acmNumber{x}
\acmArticle{x}
\acmMonth{3}

\usepackage{longtable}
\usepackage{enumitem}
\usepackage{hyperref}
\usepackage{color}
\usepackage{comment}
\usepackage{graphicx}
\usepackage{amsmath}
\usepackage{booktabs}  
\usepackage{tabularx}  
\usepackage{array}     
\usepackage{multirow}
\usepackage{xcolor}
\usepackage{longtable}
\usepackage{afterpage}
\usepackage{pifont}

\newcommand{\cmark}{\textcolor{green!70!black}{\ding{51}}}
\newcommand{\xmark}{\textcolor{red}{\ding{55}}}
\newcommand{\partially}{\textit{Partially}}

\begin{document}

\title{Networking-Aware Energy Efficiency in Agentic AI Inference: A Survey}

\authorsaddresses{Authors' Contact Information: 
Xiaojing Chen, Shanghai University, Shanghai, China, and Nanyang Technological University (NTU), Singapore; email: jodiechen@shu.edu.cn;
Haiqi Yu, Shanghai University, Shanghai, China; email: yuyuyu12123@shu.edu.cn;
Wei Ni, Edith Cowan University, Perth, Australia; email: wei.ni@ieee.org; 
Dusit Niyato, NTU, Singapore; email: dniyato@ntu.edu.sg;
Ruichen Zhang, NTU, Singapore; email: ruichen.zhang@ntu.edu.sg;
Xin Wang, Fudan University, Shanghai, China; email: xwang11@fudan.edu.cn; 
Shunqing Zhang (corresponding author), Shanghai University, Shanghai, China; email: shunqing@shu.edu.cn;
Shugong Xu, Xi'an Jiaotong-Liverpool University, Suzhou, China; email: shugong.xu@xjtlu.edu.cn.
}

\thanks{This work was supported by the National Key R\&D Program of China grants 2022YFB2902304 and 2022YFB2902303. }

\settopmatter{printacmref=false}
\setcopyright{none}
\renewcommand\footnotetextcopyrightpermission[1]{}

\author{Xiaojing Chen}
\affiliation{%
  \institution{Shanghai University, Shanghai, China, and Nanyang Technological University}
  \country{Singapore}
}
\author{Haiqi Yu}
\affiliation{%
  \institution{Shanghai University, Shanghai}
  \country{China}
}
\author{Wei Ni}
\affiliation{%
  \institution{Edith Cowan University, Perth}
  \country{Australia}
}
\author{Dusit Niyato}
\affiliation{%
  \institution{Nanyang Technological University}
  \country{Singapore}
}

\author{Ruichen Zhang}
\affiliation{%
  \institution{Nanyang Technological University}
  \country{Singapore}
}

\author{Xin Wang}
\affiliation{%
  \institution{Fudan University, Shanghai}
  \country{China}
}

\author{Shunqing Zhang}
\affiliation{%
  \institution{Shanghai University, Shanghai}
  \country{China}
}
\author{Shugong Xu}
\affiliation{%
  \institution{Xi'an Jiaotong-Liverpool  University, Suzhou}
  \country{China}
}

\renewcommand{\shortauthors}{Y. Liu et al.}
\begin{abstract}
The rapid emergence of Large Language Models (LLMs) has catalyzed \textit{Agentic artificial intelligence (AI)}, autonomous systems integrating perception, reasoning, and action into closed-loop pipelines for continuous adaptation. While unlocking transformative applications in mobile edge computing, autonomous systems, and next-generation wireless networks, this paradigm creates fundamental energy challenges through iterative inference and persistent data exchange. Unlike traditional AI where bottlenecks are computational Floating Point Operations (FLOPs), Agentic AI faces compounding computational and communication energy costs.  
In this survey, we propose an energy accounting framework identifying computational and communication costs across the Perception-Reasoning-Action cycle. We establish a unified taxonomy spanning model simplification, computation control, input and attention optimization, and hardware-aware inference. We explore cross-layer co-design strategies jointly optimizing model parameters, wireless transmissions, and edge resources. Finally, we identify open challenges of federated green learning, carbon-aware agency, 6th generation mobile communication (6G)-native Agentic AI, and self-sustaining systems, providing a roadmap for scalable autonomous intelligence.
\end{abstract}

\ccsdesc[500]{General and reference~Surveys and overviews}
\ccsdesc[500]{Computing methodologies~Artificial intelligence}
\ccsdesc[500]{Networks~Mobile networks}


\keywords{Agentic AI, energy efficiency, edge computing, mobile networks}

\maketitle
\section{Introduction}
\label{sec:intro}

The rapid emergence of Large Language Models (LLMs) and multimodal architectures has fundamentally transformed the landscape of artificial intelligence (AI), catalyzing the rise of \textit{Agentic AI}. The autonomous Agentic AI systems are capable of perceiving, reasoning, and acting in dynamic environments through a closed Perception-Reasoning-Action loop~\cite{zhang2026edge, yao2023react}. Unlike traditional AI systems that operate under static, single-turn interactions, Agentic AI integrates these capabilities into an iterative, feedback-driven pipeline: The perception module interprets multimodal sensory data (e.g., vision, audio, text) and retrieves relevant context through Retrieval-Augmented Generation (RAG); The reasoning module, powered by LLMs or multimodal foundation models, performs complex planning, causal inference, and decision-making via techniques like Chain of Thought (CoT) prompting \cite{wei2022chain}; and the action module executes decisions through tool use, Application Programming Interface (API) calls, communication, or direct control of physical actuators~\cite{shinn2024reflexion}. This architecture operates iteratively, with action outcomes fed back into perception, enabling continuous adaptation and replanning in dynamic, unstructured environments.

This paradigm shift has unlocked transformative applications across diverse domains. In mobile edge computing (MEC), Agentic AI enables context-aware services that adapt to user behavior and network conditions in real time. In autonomous driving and robotics, it supports sophisticated decision-making under uncertainty, integrating sensor fusion with high-level planning. In next-generation wireless networks, Agentic AI promises self-organizing and self-healing network management through closed-loop control~\cite{usman2025ai}. However, this closed-loop nature necessitates continuous interaction with the environment, leading to recurrent inference cycles and persistent data exchange between distributed components, thereby creating energy demands that fundamentally challenge sustainable deployment.

\subsection{Motivation}

The large-scale deployment of Agentic AI is fundamentally constrained by energy bottlenecks in inference that manifest across two interconnected dimensions: computational energy and communication energy. 

First, the massive parameter scale of modern LLMs, ranging from billions to hundreds of billions of parameters, translates directly into high computational and memory energy costs. Mechanisms such as multi-step reasoning with CoT prompting, quadratic-scaling attention operations, and large intermediate representations in multimodal pipelines substantially increase power consumption. As a result, inference has become one of the dominant contributors to overall system energy demand \cite{husom2025sustainable}. The shift from single-pass forward propagation in traditional AI to closed-loop iterative reasoning in Agentic AI changes the primary bottleneck from computation, i.e., Floating Point Operations (FLOPs), to memory bandwidth, i.e., Input/Output (I/O) Wall, exacerbating energy consumption through frequent off-chip dynamic random access memory (DRAM) accesses that consume approximately 640 pJ per access, which is orders of magnitude more expensive than arithmetic operations~\cite{sze2017efficient}.

Second, agentic systems operate in inherently distributed networks where agents continuously interact with cloud servers, edge devices, and peer agents over wired and wireless links. These interactions incur substantial communication energy overheads, including the cost of token transmission, intermediate result exchange, synchronization, and model updates across heterogeneous devices. In 6th generation mobile communication (6G) and edge scenarios, limited bandwidth, stringent latency requirements, and constrained battery capacity further magnify the energy burden of networking~\cite{zhang2025beyond, du2024enabling}. The result is a compounding effect: Computation and communication jointly form critical energy bottlenecks that must be addressed holistically to enable sustainable Agentic AI deployment.


\begin{table*}[tbp]
    \centering
    \caption{Summary of Related Works} 
    \label{tab:related_works}
    \scriptsize
    \renewcommand{\arraystretch}{1} 
    \begin{tabularx}{\textwidth}{l X c c c c} 
        \toprule
        \textbf{Ref.} & 
        \textbf{Overview} & 
        \textbf{\shortstack{Agentic\\AI}} & 
        \textbf{\shortstack{Edge\\Inference}} & 
        \textbf{\shortstack{Wireless\\Networks}}  & 
        \textbf{\shortstack{Energy-\\Efficiency}} \\ 
        \midrule

        \cite{11316910} & 
        A comparative study evaluating the functional capabilities and reasoning accuracy of general-purpose Agentic AI frameworks (e.g., AutoGPT), but lacking insights into the energy constraints or communication overhead required for real-world wireless edge deployment. & 
        \cmark & 
        \xmark & 
        \xmark & 
        \xmark \\ 
        \hline
        
        \cite{usman2025ai} & 
        A detailed article proposing an edge-native intelligence framework for 6G networks that integrates AI/ML for autonomous network management and self-healing, addressing edge deployment constraints but without a specific focus on the generative agentic workflow. & 
        \partially & 
        \cmark & 
        \cmark & 
        \xmark \\ 
        \hline
        \cite{11158258} & 
        A survey reviewing advanced quantization techniques (GPTQ, GGUF, AWQ) for Large Language Models (LLMs) within autonomous driving and ADAS, focusing on memory reduction and inference acceleration on resource-limited edge hardware while analyzing accuracy trade-offs. & 
        \xmark & 
        \cmark & 
        \xmark & 
        \cmark \\ 
        \hline
        \cite{bhardwaj2024survey} & 
        A survey delving into the integration of Large Language Models (LLMs) with edge computing, highlighting advantages in privacy and latency while addressing challenges in computational demand and energy efficiency for resource-limited device deployment. & 
        \xmark & 
        \cmark & 
        \cmark & 
        \partially \\ 
        \hline

        \cite{zhang2026edge} & 
         A comprehensive survey on Agentic AI frameworks for edge intelligence, introducing enabling technologies, representative case studies,
        and future directions toward scalable and trustworthy deployments in
        next-generation wireless edge networks. & 
        \cmark & 
        \cmark & 
        \cmark & 
        \xmark \\ 
        \hline
        
        \cite{dantas2025compression} & 
        A comprehensive review of state-of-the-art LLM compression techniques (e.g., pruning, quantization) aimed at lowering inference energy consumption, yet without addressing the communication overhead or distributed coordination inherent in wireless agentic networks. & 
        \xmark & 
        \cmark & 
        \xmark & 
        \cmark \\ 
        \hline

        \cite{liu2025adaptiveresourceefficientagenticai} & 
        A systematic survey reviewing inference optimization techniques for Agentic AI, focusing on computational model compression (e.g., quantization), with limited discussion on communication energy overhead or cross-layer synergy in wireless environments. & 
        \cmark & 
        \cmark & 
        \partially & 
        \partially \\ 
        \hline
        
        \textbf{Ours} & 
        A survey and tutorial on Green Agentic AI bridges algorithmic optimization with network constraints. It proposes a unified taxonomy spanning model simplification, computation control, and cross‑layer co‑design to enable sustainable closed‑loop inference in resource‑constrained wireless edge networks. 
        & 
        \cmark & 
        \cmark & 
        \cmark & 
        \cmark \\ 
        \bottomrule
    \end{tabularx}
\end{table*}

Table \ref{tab:related_works} summarizes the existing works. While existing surveys on LLM efficiency~\cite{dantas2025compression, 11158258} heavily prioritize algorithmic compression methods like quantization and pruning, they largely overlook the networking dimension; even recent attempts to optimize agent inference~\cite{liu2025adaptiveresourceefficientagenticai} tend to focus narrowly on computational model compression rather than adopting a broader view of cross-layer synergy. Conversely, research into edge AI and wireless networks~\cite{zhang2026edge, bhardwaj2024survey} often treats AI inference merely as a black-box workload, failing to account for unique agentic demands, such as iterative reasoning and semantic communication. Similar limitations appear in broader network integration and security studies~\cite{usman2025ai}, which address autonomous management but miss the distinct energy dynamics of the generative agentic loop; while functional evaluations~\cite{11316910} assess reasoning skills to the exclusion of real-world energy constraints. To bridge these fragmented perspectives, our survey introduces the concept of energy-efficient Agentic AI and proposes a unified taxonomy that combines model simplification with computation control and cross-layer co-design, offering actionable guidelines for sustainable deployment in resource-constrained wireless edge networks.

\subsection{Contributions}

This survey provides a systematic overview of networking-aware energy-efficient techniques for Agentic AI inference, with the following key contributions:

\begin{itemize}[leftmargin=*]
    \item We propose a comprehensive energy accounting framework that dissects the unique inference costs in Agentic AI systems, distinguishing between computational energy (arithmetic operations, memory access) and communication energy as the two primary bottlenecks in the closed-loop Perception-Reasoning-Action cycle.
    
    \item We provide a unified taxonomy of energy-efficient optimization methods, categorized into four pillars: (1) \textit{model simplification} (quantization, pruning, distillation, sparse Mixture-of-Experts (MoE) activation, action-oriented simplification); (2) \textit{computation control} (token length control, early exit/adaptive depth, layer skipping, decoding simplification, workload scheduling); (3) \textit{input and attention optimization} (token pruning, sparse attention, Key-Value (KV) caching and reuse); and (4) \textit{hardware-aware inference} (precision scheduling, Dynamic Voltage and Frequency Scaling (DVFS), memory and I/O optimization). We analyze their interrelationships, complementarities, and inherent trade-offs between energy efficiency, accuracy, and latency, with particular attention to their deployment in resource-constrained edge environments. 
    
    \item We present a detailed exploration  of cross-layer co-design strategies that jointly optimize AI model parameters, wireless transmissions, and edge computing resources. We categorize these into: cross-layer optimization variables (transmission-inference coupling, mobility-aware scheduling, model-channel adaptation); user-edge-cloud collaboration (split inference, adaptive offloading, collaborative caching); and communication-inference co-design (semantic communication, retrieval-augmented communication, energy-aware scheduling).
    
    \item We identify open challenges and future research directions spanning federated green learning, carbon-aware agency, 6G-native Agentic AI, and self-sustaining systems, providing a roadmap for achieving scalable, autonomous intelligence in resource-constrained environments.
\end{itemize}

\subsection{Paper Organization}

\begin{figure*}[ht]
    \centering
    \includegraphics[width=0.83\linewidth]{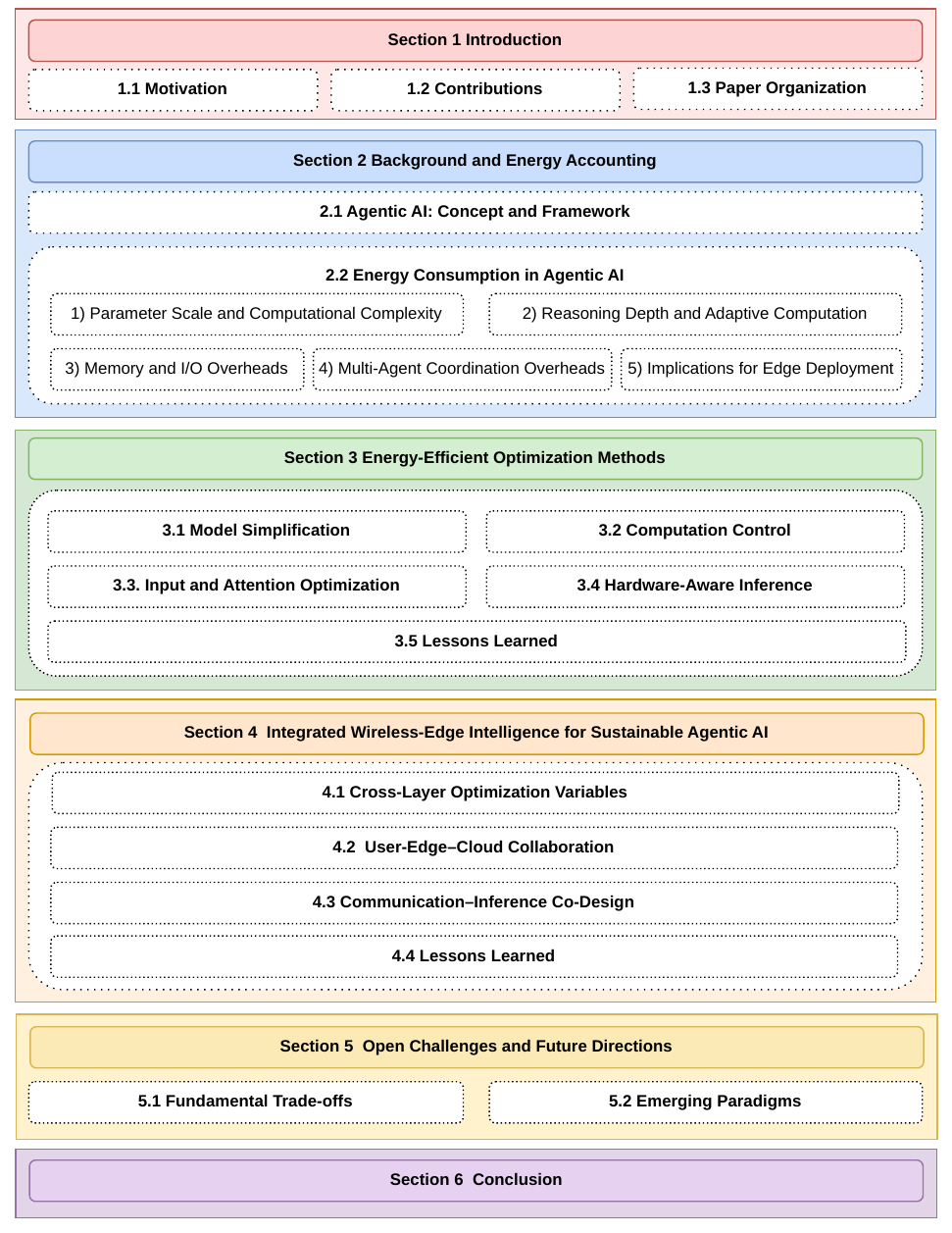} 
    \caption{Survey organization: Section~\ref{sec:background} introduces Agentic AI concepts and energy accounting; Section~\ref{sec:methods} reviews optimization methods; Section~\ref{sec:crosslayer} examines cross-layer co-design for wireless-AI integration; Section~\ref{sec:future} identifies open challenges and future directions; Section~\ref{sec:conclusion} provides concluding remarks.}
    \label{fig:paper_structure}
\end{figure*}

The remainder of this survey is organized as follows. Section~\ref{sec:background} provides the background on the Agentic AI concepts and presents a detailed energy accounting framework dissecting computational and communication costs. Section~\ref{sec:methods} reviews energy-efficient optimization methods across the four pillars of our taxonomy. Section~\ref{sec:crosslayer} examines cross-layer co-design strategies for joint wireless AI optimization. Section~\ref{sec:future} outlines open challenges and future directions. Section~\ref{sec:conclusion} concludes the survey. Fig.~\ref{fig:paper_structure} illustrates the overall structure of this survey.

\section{Background and Energy Accounting}
 \label{sec:background} 
 \subsection{Agentic AI: Concept and Framework}
\label{subsec:agentic_concept}

While the concept of autonomous agents has evolved through multiple paradigms, from rule-based systems to reinforcement learning-driven controllers, the current generation of Agentic AI represents a qualitative shift in capability and complexity. Table~\ref{tab:evolution_chain} delineates this evolutionary trajectory, contrasting four stages of intelligent agent development. Early deep reinforcement learning (DRL)-powered agents relied on policy networks, e.g., Deep Q-Network (DQN), and Proximal Policy Optimization (PPO), with implicit, weak reasoning capabilities and short-term Markov state memory, achieving very inference energy but poor generalization. The subsequent LLM-powered agent introduced static probabilistic generation with fixed context windows, enabling sophisticated text processing but lacking true autonomy due to the absence of feedback loops. 

Standard Agentic AI achieves the maximum capability through the integration of perception, reasoning, and action into a closed-loop architecture, where action outcomes dynamically inform subsequent perception and reasoning iterations. This closed-loop design enables autonomous planning with dynamic long-horizon reasoning and long-term vector database memory, but at the cost of extremely high energy consumption that grows multiplicatively with reasoning depth and agent interactions. Consequently, this survey focuses on the fourth evolutionary stage: \textit{Energy-Efficient Agentic AI}, which seeks to reconcile advanced autonomy with the strict resource constraints of edge and mobile deployment.

\begin{table*}[tbp] 
    \centering
    \caption{Evolutionary Chain of Intelligent Agents: From Specialized Optimization to Sustainable Autonomy}
    \label{tab:evolution_chain}
   \scriptsize
    \renewcommand{\arraystretch}{1}
    \begin{tabularx}{\textwidth}{l X X X X}
        \toprule
        \textbf{Dimension} & \textbf{DRL-Powered Agents} & \textbf{LLM-Powered Agents} & \textbf{Standard Agentic AI} & \textbf{Energy-Efficient Agentic AI } \\
        \midrule
        \textbf{Core Driver} & Policy Networks & Static LLMs & Perception-Reasoning-Action Loop & Resource-Aware Co-Design \\
        \textbf{Paradigm} & Trial-and-Error Learning & Probabilistic Generation & Autonomous Planning & Adaptive Computation and Communication \\
        \textbf{Reasoning} & Implicit / Weak & Static CoT & Dynamic Long-Horizon & Elastic Depth with Early Exit \\
        \textbf{Memory} & Short-term Markov State & Fixed Context Window & Long-term Vector Database & Compressed KV Cache \\
        \textbf{Energy Profile} & Very Low (Inference) & High (Linear Growth) & Extremely High (Multiplicative) & Controllable via Algorithm-Hardware Co-Design \\
        \textbf{Limitation} & Poor Generalization & Lack of Autonomy & High Cost \& Latency & Accuracy Trade-off \\
        \textbf{Use Case} & Specialized Control & Text Processing & Cloud Orchestration & Mobile/Edge Autonomy \\
        \bottomrule
    \end{tabularx}
\end{table*}

The functional architecture of Agentic AI comprises four interconnected modules, each presenting distinct energy optimization opportunities. Table~\ref{tab:energy_breakdown} characterizes the energy consumption profile across these modules, revealing distinct hardware utilization patterns that inform targeted optimization strategies.

\textbf{Perception Module.} This module processes multimodal sensory inputs, including visual data from cameras, acoustic signals from microphones, structured data from databases, and contextual information from knowledge bases, to construct a structured representation of the environment. Unlike passive sensing in traditional systems, Agentic AI perception is active and goal-directed, employing RAG to dynamically retrieve relevant external knowledge in response to task requirements~\cite{zhang2025toward}. 
As shown in Table~\ref{tab:energy_breakdown}, the Perception module is predominantly \textit{compute-bound}, with high Graphics Processing Unit (GPU) compute energy consumption driven by vision encoder FLOPs and medium memory energy from activation storage. This aligns with empirical findings that multimodal inputs incur 3.0$\times$--67.9$\times$ energy overhead compared to text-only baselines during visual encoding~\cite{moghadampanah2025energy}, motivating input compression and model simplification as critical optimizations.

\textbf{Reasoning Module.} Powered by LLMs or multimodal foundation models, this module performs high-level cognitive functions, including planning, causal inference, analogical reasoning, and decision-making under uncertainty. Techniques such as CoT prompting~\cite{wei2022chain}, tree-of-thought search, and iterative self-refinement~\cite{shinn2024reflexion} enhance reasoning quality but substantially increase computational depth and energy consumption. 
As illustrated in Table~\ref{tab:energy_breakdown}, the Reasoning module contrasts Perception: It is overwhelmingly \textit{memory-bound} with very high memory energy consumption from weight storage and KV cache management, while compute energy remains medium despite intensive arithmetic operations. This underscores why KV cache designs, adaptive computation depth, and early termination mechanisms are essential for sustainable agentic inference.

\textbf{Action Module.} This module translates high-level decisions into executable outputs, encompassing tool use (e.g., invoking calculators, search engines, or code interpreters), API calls to external services, communication with other agents or humans, and direct control of physical actuators in robotic systems ~\cite{zhang2026edge}. While individual actions may involve lightweight computation, the Action module influences system-level energy efficiency through its impact on the closed-loop iteration frequency: Poor action choices may necessitate expensive re-planning cycles, while well-calibrated actions can terminate reasoning early~\cite{hao2022multiagent}. Table~\ref{tab:energy_breakdown} reveals that the Action module exhibits a unique profile with high communication energy dominating over compute and memory costs, reflecting the overhead of multi-agent synchronization, API calls, and distributed coordination. This highlights the need for communication-efficient orchestration and edge-native deployment to minimize synchronization overhead.

\textbf{Memory Module.} Supporting the other modules, the Memory module maintains short‑term working memory (e.g., conversation history, intermediate reasoning) and long‑term episodic/semantic memory (e.g., vector databases of prior experiences and skills). Moving from fixed context windows in standard LLMs to dynamic, expandable memory in Agentic AI introduces significant bandwidth and storage energy costs, especially for long‑horizon tasks.  
As shown in Table~\ref{tab:energy_breakdown}, the Memory module is predominantly \textit{bandwidth‑bound}, with high GPU energy consumption driven by KV cache fragmentation and frequent GPU–CPU/DRAM transfers~\cite{luo2025simllm}. This motivates KV cache management and memory compression as high‑impact optimizations. Medium communication energy arises from cache synchronization across distributed agents.

\begin{table*}[t]
\caption{Energy Consumption Characteristics by Agentic AI Module}
\label{tab:energy_breakdown}
\centering
\renewcommand{\arraystretch}{1.1}
\resizebox{\textwidth}{!}{%
\begin{tabular}{@{}lccccc@{}}
\toprule
\textbf{Module} & \textbf{Primary Operation} & \textbf{Compute Energy} & \textbf{Memory Energy} & \textbf{Communication Energy} & \textbf{Key Bottleneck} \\
\midrule
\textbf{Perception} & Vision encoding, RAG retrieval & High & Medium & Low-Medium & Vision encoder FLOPs \\

\textbf{Reasoning} & LLM inference, CoT generation & Medium & Very High & Low-Medium & KV cache capacity \\

\textbf{Action} & Tool/API execution, coordination & Medium & Low & High & Multi-agent communication \\

\textbf{Memory} & KV cache management, retrieval & Low & Very High & Medium & Cache fragmentation \\
\bottomrule
\end{tabular}%
}
\end{table*}

\subsection{Energy Consumption in Agentic AI}
\label{subsec:energy_consumption}

Agentic AI inference, encompassing the complete Perception-Reasoning-Action pipeline, constitutes the primary source of energy consumption in Agentic AI systems. Unlike traditional AI inference characterized by single-pass forward propagation, Agentic AI operates through closed-loop iterative reasoning that fundamentally alters the energy consumption profile. 
This subsection analyzes the key contributors to inference energy costs and their implications for deployment.


\subsubsection{\textbf{Parameter Scale and Computational Complexity}}
The sheer size of modern LLMs, ranging from billions to hundreds of billions of parameters, directly impacts energy consumption across both the Reasoning and Perception modules. Each forward pass in the Reasoning module requires extensive matrix multiplications and attention computations, which scale quadratically with sequence length in standard transformer architectures. Consequently, inference energy grows with both model size and input length, making long-context reasoning costly. Meanwhile, the Perception module's vision encoders introduce additional computational burdens, with studies showing that visual encoding can dominate total energy consumption in multimodal agentic pipelines~\cite{moghadampanah2025energy}.

\subsubsection{\textbf{Reasoning Depth and Adaptive Computation}}
Agentic AI systems often employ multi-step reasoning, such as CoT prompting~\cite{wei2022chain,wang2023self} and iterative planning~\cite{yao2023react,shinn2024reflexion}. While these techniques improve reliability and accuracy by decomposing complex problems into manageable steps, they significantly increase inference energy by requiring multiple forward passes or deeper activation of layers. The energy cost grows multiplicatively with reasoning depth: Each reasoning step invokes the full memory-bound overhead of the Reasoning module, while potentially triggering additional Perception (for gathering new information) and Action (for executing intermediate steps) operations. Adaptive reasoning strategies that dynamically adjust depth based on task complexity, e.g., early exit mechanisms or uncertainty-aware halting, are therefore critical to balance accuracy with energy efficiency.

\subsubsection{\textbf{Memory and I/O Overheads}}  
In Agentic AI inference, energy use is driven not only by arithmetic operations but also by memory access and data movement, which form a major bottleneck. Transformer architectures rely on attention mechanisms requiring frequent KV reads and writes, while multimodal models generate large intermediate representations repeatedly transferred across memory hierarchies.
Empirical studies reveal that memory transfers can consume significantly more energy than computation. For example, accessing off-chip DRAM incurs an energy cost of approximately 640~pJ with latency of about 100~ns, whereas a standard arithmetic operation requires less than 1~pJ and completes within 1~ns~\cite{sze2017efficient,hennessy2017computer}. This disparity underscores that inference efficiency is often constrained more by memory and I/O overheads than by raw computation, highlighting the critical role of techniques, such as compressed KV cache designs, and memory-aware scheduling in reducing energy costs during Agentic AI inference. 

\subsubsection{\textbf{Multi-Agent Coordination and Communication Overheads}}
A distinctive feature of Agentic AI is the potential for multiple agents to concurrently perform inference and exchange intermediate results. This parallelism, while enabling sophisticated collaborative behaviors, amplifies energy costs through the Action module's high communication energy component. Agents repeatedly query large models for perception, reasoning, and action, while synchronizing state, sharing context, and coordinating decisions across distributed nodes.
Without careful orchestration, redundant inference across agents leads to wasted energy: Multiple agents may independently perform similar perception tasks on overlapping environmental data, or re-compute reasoning steps that could be shared. Furthermore, the communication energy for state synchronization and result exchange can dominate local computation, particularly in bandwidth-constrained wireless environments, highlighting the need for communication-efficient orchestration and edge-native deployment to minimize synchronization overhead.

\subsubsection{\textbf{Implications for Edge Deployment}}
For mobile and edge devices with limited battery and compute capacity, the compounded energy footprint of Agentic AI inference poses a critical deployment bottleneck. The multiplicative energy growth from closed-loop reasoning, where each iteration invokes compute-bound Perception, memory-bound Reasoning, communication-heavy Action, and bandwidth-bound Memory operations, rapidly exhausts available resources. High energy demand restricts continuous operation time, raises thermal concerns that trigger throttling, and limits the complexity of tasks that can be performed autonomously. 

\begin{figure}[t]
    \centering
    \includegraphics[width=0.7\linewidth]{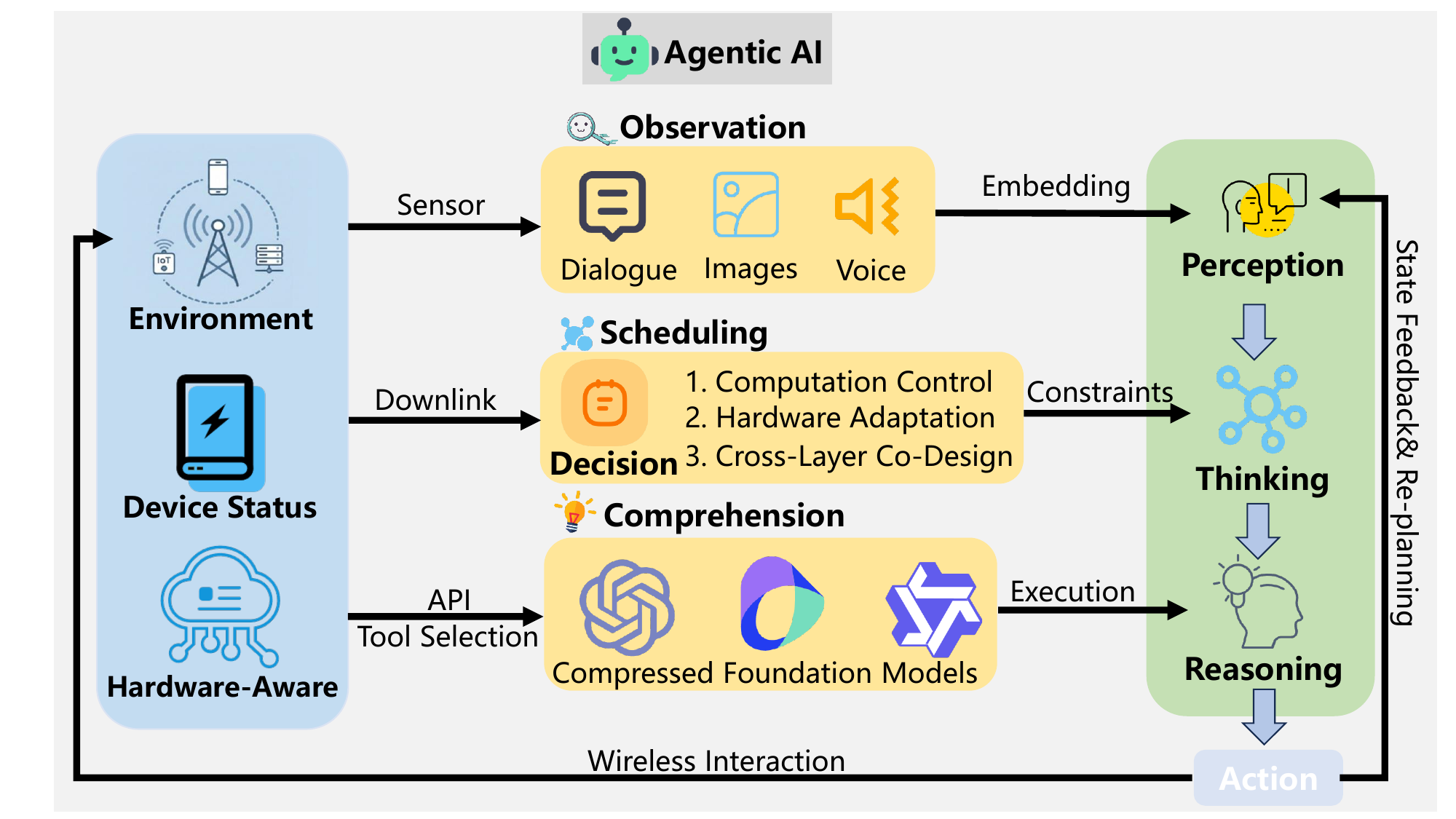} 
    \caption{Generic framework illustrating how Edge Agentic AI autonomously integrates multimodal observations, scheduling-driven constraints, and compressed foundation models with hardware-aware API selection for the continuous Perception–Thinking–Reasoning–Action process in resource-constrained wireless networks.}
    \label{fig:framework}
\end{figure}

To intuitively illustrate the coupling mechanism between these energy factors and the deployment environment, Fig.~\ref{fig:framework} presents the generic framework of Agentic AI. As shown in this figure, the agent is situated within a wireless network environment, processing multimodal input contexts, including dialogue, images, and speech, and converting them into high-dimensional embedding vectors to drive the core Perception-Reasoning-Action loop. This closed-loop process is strictly governed by scheduling decisions, which directly reflect the previously analyzed energy bottlenecks and constitute dynamic constraints on the agent's behavior. This mechanism ensures that the agent can leverage underlying inference frameworks to optimize computation paths under resource-constrained conditions, efficiently performing reasoning and wireless interactions.

In summary, the energy consumption of Agentic AI inference arises from the interplay of parameter scale, reasoning depth, memory/I/O overheads, and multi-agent coordination across the four core modules. Addressing these challenges requires a combination of algorithmic, architectural, and system-level optimizations to ensure that agentic intelligence can operate efficiently in real-world, energy-constrained scenarios. 

\section{Energy-Efficient Optimization Methods}
\label{sec:methods}

Energy‑efficient optimization methods are central to reducing the computational and communication overheads of Agentic AI inference. With massive parameter scales in LLMs and multimodal models, and the distributed nature of agentic systems, energy costs arise not only from arithmetic operations but also from memory transfers, networking, and redundant multi‑agent execution~\cite{biswas2024power,ieee2024pruning}. To address these challenges, diverse techniques have emerged, including model simplification, computation control, input and attention optimization, and hardware‑aware inference~\cite{dantas2025compression,bullo2024dynamic,liu2024moe}. These methods form a unified taxonomy that balances energy efficiency, accuracy, and latency, enabling sustainable deployment of Agentic AI across heterogeneous edge–cloud networks.

\subsection{Model Simplification}
\label{sec:model-simp}
Model simplification reduces the structural complexity of large models while preserving functionality, making them lighter, faster, and more energy‑efficient without major accuracy loss~\cite{dettmers2023qlora,ray2024llmedge}. Techniques include quantization, pruning, distillation, architectural redesign, and simplifying action spaces.  
For Agentic AI inference, simplification is vital since billion-parameter models demand extensive memory bandwidth, driving high energy use in computation and data movement~\cite{zhu2024compression}. Smaller or pruned models cut floating-point operations and processor power~\cite{an2024flap}. Techniques such as quantization reduce memory usage and limit DRAM access, one of the most energy-intensive operations~\cite{sze2017efficient}. Lower latency accelerates inference, allowing devices to return to low-power states. Reduced size and energy demand also enable deployment on mobile and Internet of Things (IoT) platforms, supporting sustainable operation in constrained networks. In distributed or federated settings, simplified models lessen transmission overhead, improving communication efficiency and saving energy in wireless inference.

\begin{figure}[t]
    \centering
\includegraphics[width=0.7\textwidth]{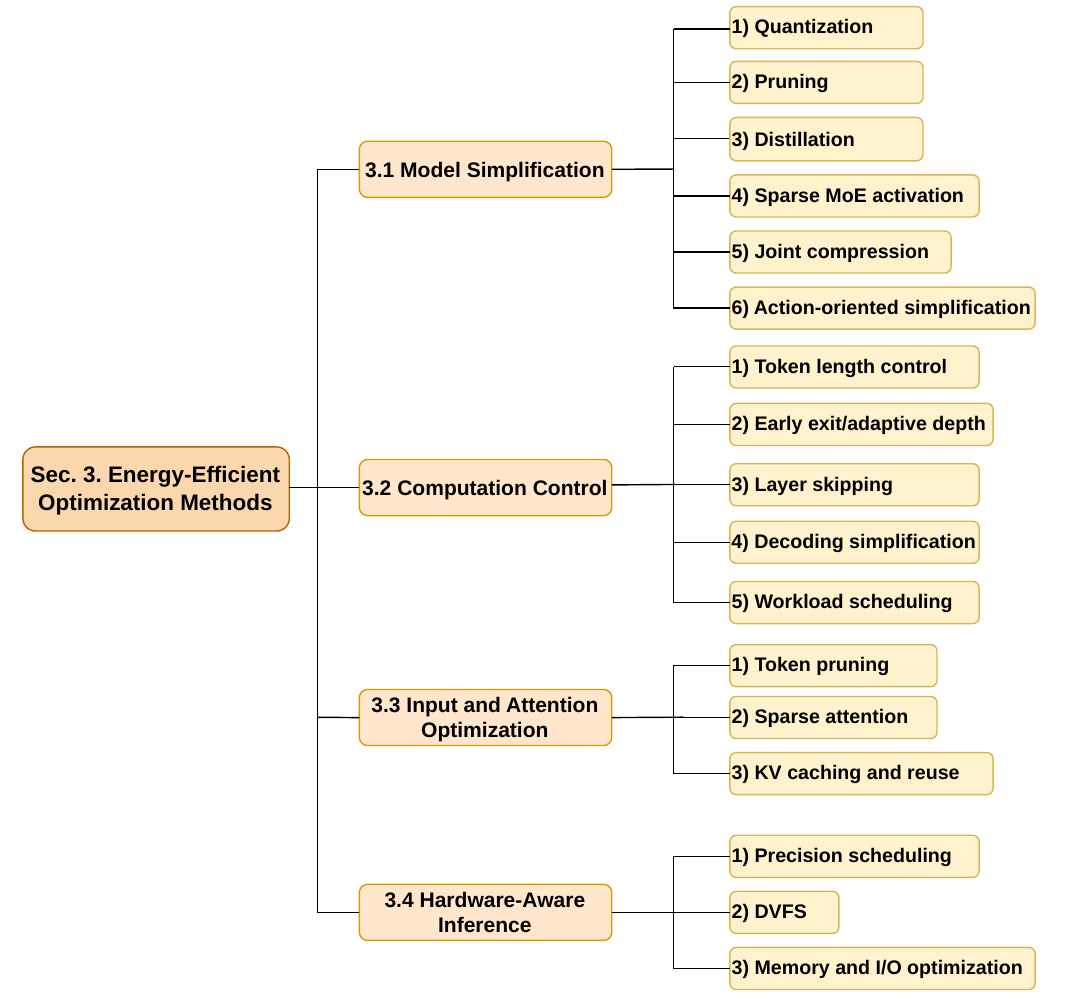} 
    \caption{An overview of energy efficient optimization methods.}
    \label{fig:sec3}
    \vspace{-13pt}
\end{figure}


\subsubsection{\textbf{Quantization}}
Quantization is a fundamental technique for simplifying large models by reducing the precision of weights and activations. Instead of storing and computing with full-precision floating-point numbers, parameters can be represented in lower bit-width formats, such as 8-bit, 4-bit, and even 3-bit~\cite{zhao2024atom}. This simplification directly reduces the size of the model, lowers the number of bits processed per operation, and enables more efficient use of hardware resources. By shrinking the computational footprint, quantization not only accelerates inference but also reduces the overall energy demand, making continuous operation more feasible in networks with limited battery and compute capacity.

Recent advances in model quantization explore extreme low‑bit representations, activation‑aware schemes, and hybrid compression strategies, collectively reducing memory, latency, and energy consumption on mobile/edge devices while balancing efficiency gains with accuracy trade‑offs. 
Husom \textit{et al.}~\cite{husom2025sustainable} evaluate 28 quantized LLMs, showing that 3-bit and 4-bit quantization can cut energy use by up to 79\% versus FP16. However, efficiency comes at an accuracy cost: Commonsense QA tasks retain near-baseline performance (accuracy drop $<$5\%), while complex reasoning and math tasks (e.g., GSM8K) degrade by 20\%--30\%. This underscores the trade-off between sustainability and predictive performance in edge deployment.
Hu \textit{et al.} \cite{hu2025qllms} propose QLLMS, a quantization‑adaptive scheduling framework for LLMs in partially informed edge systems. It jointly selects quantization levels and allocates heterogeneous resources using an available
quantization set profiler, low‑rank reconstruction, and a stable matching scheduler. 
Zhang \textit{et al.}~\cite{zhang2025beyond} propose an edge inference framework for generative LLMs in wireless networks. Incorporating quantization, batching, and communication resource allocation, the framework enables efficient deployment of transformer LLMs on constrained edge nodes. Results show quantization greatly reduces energy and computation costs, supporting real-time, privacy-preserving LLM services for mobile users.

By lowering the bit-width of weights and activations, these quantization methods decrease the size of parameters and intermediate results, enabling faster inference and lower energy consumption for Agentic AI. 

\subsubsection{\textbf{Pruning}}  

Pruning reduces the size and complexity of large models by removing redundant parameters, neurons, or entire structural components, such as attention heads and feed-forward blocks. It can be applied in either an unstructured manner (removing individual weights based on importance) or a structured manner (removing entire groups of parameters)~\cite{zhu2024compression,gao2024displlm,ling2024slimgpt}. 
By eliminating unnecessary computation, pruning helps align the model capacity with task requirements, enabling sustainable operation in distributed agentic systems.  

Recent pruning approaches demonstrate that pruning can substantially reduce model size, latency, and energy consumption of mobile/edge devices while retaining up to 95\% of baseline accuracy~\cite{ma2023llmpruner}, though smaller models and complex tasks reveal clearer trade‑offs between efficiency and performance.  
Xia \textit{et al.}~\cite{xia2024sheared} present Sheared LLaMA, an end-to-end pruning strategy that compresses standard 7B models into smaller 1.3B or 2.7B variants without extensive retraining. This approach democratizes LLM capabilities for mobile devices by aligning the model size with strict edge memory and latency budgets, underscoring pruning’s role in enabling practical deployment under energy constraints.
Frantar \textit{et al.}~\cite{frantar2023sparsegpt} introduce SparseGPT, a one-shot pruning method achieving 50--60\% sparsity in massive GPT models (e.g., OPT-175B, BLOOM-176B) within 4.5 hours. Despite pruning over 100B weights, accuracy loss is minimal: At 50\% sparsity, OPT-175B perplexity rises only slightly (8.35 to 8.21), while smaller OPT-1.3B increases from 14.62 to 17.46. These results highlight the efficiency--accuracy trade-off: Large models tolerate high sparsity with negligible degradation, whereas smaller ones show more noticeable drops.
Tian \textit{et al.}~\cite{greenllm2024} propose GreenLLM, an energy-aware pruning framework for edge LLMs. Guided by hardware energy estimation and space–weight–power (SWaP) constraints, a generative pruning-ratio model and dependency-aware pruner preserve key abilities, while low-rank adaptation (LoRA) fine-tuning restores performance. On Llama-7B, GreenLLM cuts energy and latency by over 30\% with acceptable perplexity and accuracy, underscoring its value for sustainable edge inference.


Pruning reduces redundant computation and memory usage, lowering energy demand and enabling deployment on resource‑constrained devices, while strengthening Agentic AI by supporting more responsive perception and faster reasoning. 

\subsubsection{\textbf{Distillation}}  

Distillation is a model compression technique that transfers knowledge from a large teacher model into a smaller student model \cite{gu2024minillm}. The goal is to preserve the teacher's capabilities while reducing the computational and memory footprint of the student, thereby lowering inference energy consumption. For Agentic AI inference, distillation is valuable because it enables lightweight models to perform complex reasoning and perception tasks with reduced energy demand. 

Recent distillation techniques for LLMs demonstrate how knowledge distillation enables energy‑\break efficient, low‑latency deployment of compact student models across networks.
Liu \cite{liu2024aicd} introduces Autoregressive In‑Context Distillation (AICD), a unified objective to distill both in‑context learning and reasoning abilities from large teacher LLMs into smaller student models. AICD applies meta‑teacher forcing on CoT exemplars and leverages the autoregressive nature of LLMs to jointly optimize the likelihood of all rationales in one pass.
Liu \textit{et al.}~\cite{liu2024mobilellm} propose MobileLLM, a sub‑billion parameter framework tailored for smartphones. Using “deep and thin” architectures with progressive knowledge transfer, it outperforms 125M/350M baselines by 2.7\%/4.3\% on zero‑shot reasoning. MobileLLM also cuts energy use: A 350M 8‑bit model consumes only 0.035 J/token versus 0.7 J/token for a 7B LLaMA‑v2, enabling all‑day on‑device reasoning under strict memory and energy budgets.
Zheng \textit{et al.} \cite{zheng2025uav} propose a dynamic knowledge distilled radio frequency fingerprints-based LLM for Unmanned Aerial Vehicle (UAV) identification in integrated sensing and communication (ISAC) networks. Using PPO to adjust distillation temperature, knowledge is transferred from a GPT‑2‑based teacher to a lightweight Lite‑HRNet student. The distilled model achieves 98.38\% accuracy with only 0.15M parameters.

These distillation methods enable energy‑efficient Agentic AI inference by compressing large models into lightweight students that require fewer computational and communication resources, thereby enhancing autonomy and proactivity across the core functions of perception, reasoning, and action in resource‑constrained networks.

\subsubsection{\textbf{Sparse MoE activation}}  

Sparse MoE activation reduces inference cost by selectively activating only a small subset of experts for each input token, rather than using all experts in a dense fashion~\cite{rajbhandari2022deepspeedmoe}. By routing tokens to specialized experts, sparse MoE improves efficiency while maintaining model capacity, making it particularly suitable for large-scale deployment in edge and mobile networks where energy budgets are limited.  

Recent innovations in MoE inference show that adaptive expert activation and deployment strategies can significantly reduce memory, latency, and energy costs while preserving accuracy, enabling practical deployment across mobile, industrial, and large‑scale networks. 
Yi \textit{et al.}~\cite{yi2025edgemoe} propose EdgeMoE, an inference engine for mobile and edge devices. By storing frequent non-expert weights in memory and dynamically loading expert weights only when needed, EdgeMoE reduces memory and runtime. With expert-wise bit-width adaptation and predictive preloading, it achieves up to 2.77$\times$ speedup and deployment of $>$10B-parameter MoE models with minimal accuracy loss ($<$2\%), showing energy savings for edge inference.  
Kong \textit{et al.}~\cite{kong2024swapmoe} introduce SwapMoE, which keeps a small set of Virtual Experts in memory mapped to original experts. Exploiting activation locality, SwapMoE cuts memory use by 67\% and latency by 50\%, with only a slight Rouge-2 drop (0.041) on summarization tasks. This balance of efficiency and accuracy makes large MoE models practical for consumer devices under strict energy and memory constraints.
Ren \textit{et al.}~\cite{11161045} address the challenge of deploying MoE
LLMs cost-effectively within edge computing networks. They formulate expert model deployment as an optimization problem, minimizing communication, computation, and storage costs. Their two-stage method strategically places experts across edge nodes, achieving significant reductions in overall deployment energy and cost.  

By selectively activating relevant experts, sparse MoE reduces computation, memory, and communication overheads, enabling scalable deployment of large models in energy‑constrained networks while enhancing Agentic AI capabilities.

\subsubsection{\textbf{Joint compression}}
Joint optimization approaches that integrate quantization, pruning, and distillation demonstrate how combining compression techniques can substantially reduce model size, memory, and energy costs while preserving accuracy, enabling efficient real‑time deployment of LLMs on resource‑constrained edge networks. 
Ahtasam \textit{et al.}~\cite{ahtasam2025dol} propose DOL‑LLM, which integrates quantization, pruning, and distillation with domain‑specific training for edge deployment. Mixed‑precision quantization cuts memory bandwidth by 4$\times$ with $<$2\% accuracy loss, while structured pruning removes 30\%–40\% of non‑critical parameters yet preserves $>$98\% functionality. Distillation compresses models to $<$25\% of original size while maintaining strong generative performance. Benchmarks show 5.7\,GB GPU RAM use, 2–4$\times$ latency speedups, and 30\%–50\% energy savings on ARM processors, highlighting the efficiency–accuracy trade‑off of domain‑oriented lightweight LLMs.
Agrawal \textit{et al.}~\cite{agrawal2025efficient} integrate pruning, quantization, and distillation to improve LLM efficiency on edge devices. Their approach reduces model size by up to 60\% and memory footprint by 50\%, while distillation retains 85\% of teacher accuracy. This collective integration enables real-time applications on constrained hardware.
Yu \textit{et al.}~\cite{yu2024edge} propose Edge-LLM with Layer-wise Unified Compression (LUC), selecting pruning and quantization policies by layer sensitivity. Coupled with adaptive tuning, this reduces memory overhead and computation depth, delivering 0.70\%–1.29\% higher accuracy than baselines under equal resource limits. Thus, pruning combined with compression significantly lowers energy use while maintaining accuracy.
 
\subsubsection{\textbf{Action-oriented simplification}}
\label{sec:action_simplification}
Beyond LLM‑centric simplification for perception and reasoning, recent work targets the action space, action models, and federated optimization to reduce Agentic AI energy use by streamlining inference and execution.  

Action space reduction narrows candidate actions to only relevant options, combining pruning and quantization: Pruning removes redundant actions, while quantization discretizes vectors into low‑bit codes. Liu \textit{et al.}~\cite{liu2024actionpruning} propose ``eSpark'', a zero‑shot framework that prunes irrelevant actions in Multi-Agent Reinforcement Learning (MARL) via LLM‑generated Python functions, refined by evolutionary search and policy feedback. Coupled with Independent PPO, eSpark improves profit/cost by up to 39\% on inventory and traffic tasks. Luo \textit{et al.}~\cite{luo2023saqrl} introduce State‑conditioned Action Quantization (SAQ), which uses a Vector Quantized Variational
AutoEncoder (VQ-VAE) to learn state‑dependent codes, avoiding exponential discretization while enforcing policy constraints. These methods cut computation, storage, and exploration costs while preserving task effectiveness under accuracy, latency, and energy constraints.

Another line of work replaces LLMs with lightweight control models in the action stage. Instead of relying on LLMs to generate instructions, compact models execute control and decision tasks with far lower computational and memory costs. 
For example, \cite{salem2025tinyml} proposes a TinyML tabular Q-learning framework for on-device control that converges in under 100 ms on low-cost microcontrollers, while Baek \textit{et al.}~\cite{baek2025slimmable} present SlimFRL, combining slimmable neural networks with federated reinforcement learning (RL) to adaptively reduce computation and communication overhead in wireless caching scenarios.
The lightweight action models not only improve deployability of Agentic AI but also provide a new pathway for energy savings in the action stage.  

Finally, federated distillation techniques integrate model compression directly into Agentic AI action, enabling compact models to be distilled and shared across devices. Ahn \textit{et al.}~\cite{ahn2020wireless} propose Wireless Federated Distillation (WFD), which combines over-the-air computation with distillation. By transmitting analog logits directly, WFD turns channel interference into constructive aggregation, reducing latency and communication energy while scaling efficiently across many devices. PFedKD is proposed in~\cite{li2025pfedkd} to advance federated distillation by tailoring models to heterogeneous IoT data. Instead of transmitting full parameters, clients share logits and prototypes distilled from a small public pseudo‑dataset, which reduces communication overhead while preserving privacy. Sharpness‑aware minimization further improves generalization, and adaptive weighting based on sample quality refines knowledge aggregation. 


These action‑oriented simplification strategies extend beyond perception and reasoning to strengthen Agentic AI’s autonomy and proactivity. By optimizing the action space through reduction, deploying lightweight control models, and enabling federated distillation, these strategies reduce energy costs while enhancing the efficiency of coordinated action across heterogeneous environments.

\begin{table*}[tbp]
    \centering
    \caption{Summary of Model Simplification Techniques (Section~\ref{sec:model-simp})}
    \label{tab:model_simplification_split}
    
    \scriptsize 
\setlength{\tabcolsep}{2pt} 
    \begin{tabularx}{\textwidth}{l|c|X|X|X}
        \hline
        \multicolumn{1}{c|}{\textbf{Technique}} & \multicolumn{1}{c|}{\textbf{Ref.}} & \multicolumn{1}{c|}{\textbf{Energy Saving Mechanism}} & \multicolumn{1}{c|}{\textbf{Benefits}} & \multicolumn{1}{c}{\textbf{Limitation}} \\ \hline

        \multirow{6}{*}{\textbf{Quantization}} 
        & \cite{husom2025sustainable} & Reduces precision to 3-bit or 4-bit formats. & Reduces energy consumption by up to 79\% compared to FP16. & Accuracy drops 20-30\% on complex reasoning tasks (e.g., GSM8K). \\ \cline{2-5}
        
        
        
        & \cite{hu2025qllms} & Quantization-adaptive resource scheduling (QLLMS). & Reduces GPU rental costs by 22.36\%; improves task completion by 59\%. & Optimization complexity increases with resource heterogeneity. \\ \cline{2-5}
        
        & \cite{zhang2025beyond} & Joint batching and quantization allocation. & Enables real-time, privacy-preserving LLM services on edge. & Resource allocation becomes NP-hard with many users. \\ \hline
        

        \multirow{4}{*}{\textbf{Pruning}} 
        & \cite{xia2024sheared} & End-to-end structured pruning (Sheared LLaMA). & Compresses 7B models to 1.3B/2.7B to fit edge budgets. & Requires retraining/fine-tuning to recover performance. \\ \cline{2-5}
        
        & \cite{frantar2023sparsegpt} & One-shot pruning (SparseGPT) for massive models. & Achieves 50-60\% sparsity without extensive retraining. & Smaller models (e.g., 1.3B) incur noticeable accuracy drops. \\ \cline{2-5}
        
        
        & \cite{greenllm2024} & Generative pruning-ratio model guided by SWaP constraints. & Reduces energy and latency by over 30\% with stable perplexity. & Dependency-aware pruning adds pipeline complexity. \\ \hline

        \multirow{5}{*}{\textbf{Distillation}} 
        
        & \cite{liu2024aicd} & Autoregressive In-Context Distillation (AICD) for CoT. & Jointly optimizes likelihood of all rationales in one pass. & Limited by the teacher's reasoning quality. \\ \cline{2-5}
        
        & \cite{liu2024mobilellm} & ``Deep and thin'' architecture design (MobileLLM). & Consumes only 0.035 J/token (vs 0.7 J for 7B); all-day usage. & Requires specific architectural redesign from scratch. \\ \cline{2-5}
        
        
        & \cite{zheng2025uav} & Distilling GPT-2 into Lite-HRNet via PPO. & Achieves 98.38\% accuracy with only 0.15M parameters. & Limited generalization beyond RF fingerprinting. \\ \hline

        \multirow{5}{*}{\shortstack[l]{\textbf{Sparse}\\\textbf{MoE}}} 
        & \cite{yi2025edgemoe} & Dynamic loading of expert weights only when activated. & Achieves up to 2.77$\times$ speedup; enables $>$10B model deployment. & I/O latency overhead during expert loading. \\ \cline{2-5}
        
        
        & \cite{kong2024swapmoe} & Swapping Virtual Experts from storage to memory. & Reduces memory by 67\% and latency by 50\%. & Minor accuracy drop (Rouge-2 score -0.041). \\ \cline{2-5}
        
        
        & \cite{11161045} & Strategic placement of experts across edge nodes. & Significant reductions in deployment energy and storage costs. & Network congestion can affect expert access. \\ \hline

        \multirow{3}{*}{\shortstack[l]{\textbf{Joint}\\\textbf{Compression}}} 
        & \cite{ahtasam2025dol} & Integrated Quantization, Pruning, and Distillation. & Compresses to $<$25\% size; 30-50\% energy savings on ARM. & Complex multi-stage optimization pipeline. \\ \cline{2-5}
        
        & \cite{agrawal2025efficient} & Joint application of pruning, quantization, distillation. & Reduces model size by 60\% and memory footprint by 50\%. & Cumulative accuracy loss from multiple compressions. \\ \cline{2-5}
        
        & \cite{yu2024edge} & Layer-wise Unified Compression (LUC) based on sensitivity. & Higher accuracy (+1.29\%) than baselines under same constraints. & Requires extensive sensitivity analysis per layer. \\ \hline

        \multirow{6}{*}{\shortstack[l]{\textbf{Action-Oriented}\\\textbf{Simplification}}} 
        & \cite{liu2024actionpruning}  & LLM generates pruning masks to remove irrelevant actions. & Increases profit/cost metrics by 39\% in MARL tasks. & Iterative evolutionary search cost during setup. \\ \cline{2-5}
        
        & \cite{luo2023saqrl} & VQ-VAE to learn state-dependent discrete action codes. & Avoids exponential blowup; enforces exact policy constraints. & Training complexity of VQ-VAE. \\ \cline{2-5}
        
        & \cite{salem2025tinyml} & Lightweight Tabular/Fuzzy Q-learning instead of Deep RL. & Fast convergence ($<$100ms); runs on low-cost MCUs. & Limited to simple control tasks (e.g., lighting). \\ \cline{2-5}
        
        & \cite{baek2025slimmable} & Adaptively adjusting neural network widths (SlimFRL). & Superior energy efficiency and robust caching performance. & Complexity in managing dynamic width adjustments. \\ \cline{2-5}        
        & \cite{ahn2020wireless} & Wireless Federated Distillation (WFD) transmitting analog logits. & Turns interference into aggregation to reduce latency and communication energy. & Susceptible to channel noise and synchronization errors. \\\cline{2-5} 
         & \cite{li2025pfedkd} & Uses pseudo-data with logits exchange; sharpness-aware minimization for generalization. & Personalized models under heterogeneous IoT data; reduced communication overhead. & Relies on pseudo-data quality; requires careful aggregation weight design. \\ \hline

    \end{tabularx}
\end{table*}

\subsection{Computation Control}  
\label{sec:computation_control}

Computation control dynamically regulates workload in Agentic AI inference, avoiding uniform execution of all layers, tokens, or decoding steps~\cite{delcorro2023skipdecode,wen2024tsmllm}. By adapting to input complexity, confidence, or resource limits, it reduces redundant operations and allocates compute where most beneficial.  
For energy efficiency, computation control cuts FLOPs, memory traffic, and latency via token length control, early exit, and selective layer skipping. Decoding simplification further improves throughput~\cite{qin2025dsbd}. Dynamic scheduling across heterogeneous hardware balances resources, avoiding over-provisioning and reducing communication or cooling costs~\cite{stojkovic2025tapas,jain2025llmbalancer}. These strategies enable scalable, energy-aware deployment of Agentic AI in constrained networks.

\subsubsection{\textbf{Token length control}}  
Token length control regulates the number of tokens generated or transmitted during inference~\cite{foster2024tokenagnostic}. In Agentic AI systems, long or redundant sequences increase computation, memory access, and communication overhead, amplifying energy use. By constraining or compressing token length, models reduce floating‑point operations, shorten latency, and lower transmission costs across distributed networks~\cite{wei2025tokencommunicationeralarge}. This makes token length control a critical strategy for energy‑efficient inference, ensuring semantic completeness while avoiding unnecessary energy expenditure.

Recent advances in token‑level optimization show that managing tokens as computational and communication units reduces latency, bandwidth, and energy while preserving accuracy in Agentic AI inference.  
Wei \textit{et al.}~\cite{wei2025tokencommunicationeralarge} propose UniToCom, which treats tokens as fundamental units for both computation and transmission. Leveraging the Generative Information Bottleneck (GenIB), their method learns concise yet informative token representations, reducing complexity and communication energy with minimal accuracy loss. They further introduce $\sigma$‑GenIB to stabilize autoregressive modeling, preserving diversity while improving efficiency.
Zhang \textit{et al.}~\cite{Zhang_2025} design a task‑oriented multimodal token transmission scheme to mitigate bandwidth and latency overhead. Using sliding‑window pooling and a weighted‑sum optimization, they jointly optimize bandwidth, power, and compressed token length, achieving efficient communication and reduced energy demand in multiuser networks.  
Yang \textit{et al.}~\cite{10778367} analyze token length control via an M/G/1 queueing model, showing that optimal token limits reduce delay and drop rates. They also propose bulk queue models for batched inference, demonstrating that managing maximum token size minimizes latency and improves energy efficiency by avoiding excessive computation and transmission.

By limiting or compressing token outputs, these methods reduce computation, memory, and communication overheads, enabling faster reasoning and sustainable deployment of Agentic AI in networks. The resulting shorter command packets further slash actuation latency and radio energy, delivering swifter, energy-aware action without sacrificing task accuracy.

\subsubsection{\textbf{Early exit/adaptive depth}}  

Early exit and adaptive depth allow language models to terminate inference once sufficient confidence is reached, rather than executing all layers for every input. The principle is that ``easy'' inputs can be processed with shallow computation, while ``hard'' cases require deeper reasoning~\cite{zeng2024consistentee}. For edge and distributed networks, early exit mechanisms are valuable because they enable lightweight devices to handle most tasks locally, while offloading complex cases to powerful servers, thus balancing accuracy with energy efficiency~\cite{11169709}.  

Recent studies illustrate diverse strategies and benefits of early exit and adaptive depth.    
Jin \textit{et al.}~\cite{11169709} introduce CE-COLLM, which partitions LLMs into lightweight edge components with early exit points and full cloud models. High-confidence tokens are generated locally, while low-confidence cases are offloaded to the cloud. This reduces communication overhead and energy consumption by avoiding unnecessary full-model inference. 
Zheng \textit{et al.}~\cite{zheng2025cloud} study cloud--edge--end collaborative inference under strict latency and resource limits. Their framework integrates early exits into hierarchical inference across devices, edge nodes, and cloud servers, terminating once confidence is sufficient to avoid redundant computation and communication. Combined with task offloading and pruning, early exit enables dynamic paths, completing simple tasks locally while escalating complex queries to the cloud.  
Venkatesha \textit{et al.}~\cite{venkatesha2025fast} propose a distributed inference framework with a lightweight draft model on edge devices and a large target model in the cloud. Early exits in the target model yield verified tokens mid-verification, allowing clients to draft subsequent tokens and reduce idle time. Deployment on the Unitree Go2 robot achieves a 21\% speedup in vision--language control, showing potential for real-time LLM/VLM applications on constrained edge devices.


By dynamically adjusting computation based on input difficulty, these methods reduce unnecessary operations and communication, enabling sustainable deployment in edge and distributed networks, while enhancing Agentic AI with faster reasoning and more proactive action.

\subsubsection{\textbf{Layer skipping}}  

Layer skipping bypasses selected transformer layers during inference. Rather than executing all blocks for each token, the model activates only layers relevant to input complexity or confidence~\cite{jiang2024dllm}. This allows lightweight devices to save energy while sustaining accuracy, adapting computation depth to task difficulty.

Diverse layer‑skipping strategies demonstrate how selectively executing or bypassing layers reduces computation, communication, and energy while preserving accuracy and resilience in Agentic AI deployments. 
Perelló \textit{et al.}~\cite{10773726} introduce JARVIS, a distributed LLM framework that splits layers across edge devices and employs skipping and recovery mechanisms. Their token-ring topology ensures robustness under node failures, demonstrating that skipping can enhance energy efficiency and resilience in distributed deployments. Larger models, such as Gemma 7B, show greater tolerance to skipping, suggesting that overparameterization improves energy-efficient adaptability.  
Abdul Hannan \textit{et al.}~\cite{11204255} present IDLD, which uses input features to determine the optimal encoder layers to execute or skip in speech foundation models. This plug-and-play mechanism outperforms random dropping and achieves comparable results to early exit, enabling efficient energy-performance trade-offs in audio applications, such as automatic speech recognition.  
Zhang \textit{et al.}~\cite{11166029} propose layer-skipping federated learning, which freezes lower layers and selectively trains upper layers. This reduces communication costs by 70\% while maintaining performance within 2\% of centralized training, highlighting how skipping can save both computation and communication energy in distributed healthcare NLP tasks.  

Layer skipping enables energy-efficient Agentic AI inference by dynamically adjusting computation depth to input complexity. It reduces redundant operations and communication overhead, making large models more practical for edge and federated networks.

\subsubsection{\textbf{Decoding simplification}}  

Decoding simplification refers to methods that reduce the computational and memory overhead of generating tokens during inference. By adopting lightweight decoding strategies or speculative mechanisms, systems can achieve faster generation with lower energy demand while maintaining acceptable output quality \cite{qin2025dsbd}.  

Speculative decoding frameworks demonstrate how parallelism and dynamic optimization reduce latency, memory, and energy while sustaining accuracy for large‑scale Agentic AI deployment on edge and mobile systems.
Xu \textit{et al.}~\cite{10812936} propose EdgeLLM, which integrates speculative decoding with width-adaptive token trees, fallback strategies, and provisional generation pipelines. This design prevents wasteful resource allocation and enables parallelism between draft and verification phases, achieving up to 9.3$\times$ speedup in per-token generation without sacrificing accuracy and reducing energy costs for large models deployed on memory-limited devices.  
Ning \textit{et al.}~\cite{ning2025dssd} present Distributed Split Speculative Decoding (DSSD), which partitions verification between edge devices and base stations. By splitting ``Accept/Reject'' decisions at the edge and ``Resample'' operations on the device, DSSD reduces communication latency and uplink transmission costs, achieving 1.5$\times$–2.4$\times$ speedups compared to conventional distributed speculative decoding. 
Zhao \textit{et al.}~\cite{zhao2024edge} propose an edge–terminal cooperative LLM framework using speculative decoding, where a small terminal LLM generates tokens and a larger edge LLM verifies them in parallel. This hybrid design lowers delay and energy versus edge‑only or terminal‑only baselines. Simulations show 25\%–34\% reductions under varying channel and device conditions, highlighting the efficiency of edge–terminal collaboration.

By reducing redundant computation and communication during token generation, these methods cut overheads and enable faster, energy‑efficient Agentic AI with sharper reasoning and proactive action in edge and distributed networks.

\subsubsection{\textbf{Workload scheduling}}  

Workload scheduling allocates inference tasks across heterogeneous resources (CPUs, GPUs, edge, cloud) to balance performance, latency, and energy~\cite{wilkins2024hybrid}. In Agentic AI, workloads vary with token length, task complexity, and deployment networks. Without efficient scheduling, systems risk idle energy waste, datacenter overheating, or excessive communication overhead. By routing requests, adapting batch sizes, and coordinating computation across devices, scheduling reduces energy demand while maintaining service quality~\cite{alizadeh2024duollm,yang2025quality}.

In response to these challenges, recent work has explored multiple scheduling strategies that directly target energy-efficient inference across heterogeneous systems. Wilkins \textit{et al.}~\cite{wilkins2024hybrid} propose a workload-aware hybrid framework that dynamically allocates queries to CPUs or GPUs depending on token length. This reduces overall CPU+GPU energy consumption by 7.5\% compared to workload-unaware baselines.  
Stojkovic \textit{et al.}~\cite{stojkovic2025tapas} introduce TAPAS for LLM inference in cloud datacenter, which leverages historical telemetry to optimize GPU virtual machine (VM) placement and workload routing under cooling and power constraints. 
These works highlight how thermal-aware scheduling balances energy savings with high-performance serving. 

Extending beyond datacenter-focused solutions, emerging research emphasizes energy‑aware scheduling frameworks tailored for both edge and cloud deployments, showcasing how adaptive routing and decentralized optimization can sustain efficiency under diverse operating conditions.
Alizadeh \textit{et al.}~\cite{alizadeh2024duollm} present Duo-LLM, which integrates auxiliary modules to enable dynamic token routing based on task complexity. This adaptive scheduling ensures that computational resources are allocated efficiently, reducing unnecessary energy use for simple tasks.  
Zhang \textit{et al.}~\cite{10571127} investigate batching and quantization-aware scheduling for large LLMs on resource-constrained edge devices. By prioritizing requests with shorter output lengths, their framework minimizes latency violations and memory footprint, hence improving throughput under strict energy budgets.  
Habibi \textit{et al.}~\cite{11095716} propose a distributed edge inference framework with fair cost-efficient incentive mechanism and adaptive dynamic scheduling algorithm . Using auction-based device selection and deadline-driven scheduling, the system cuts communication overhead by 54.7\%, optimizing pipeline-parallel inference under strict constraints.
Li \textit{et al.}~\cite{li2024collm} propose CoLLM, a collaborative LLM inference framework for edge devices. By distributing attention/MLP via tensor parallelism and balancing workloads with latency‑aware partitioning, CoLLM achieves 1.9$\times$–2.3$\times$ speedup over hierarchical methods. Energy tests show 1,000~MB transmission costs 120–150 mAh, while four‑device collaboration reduces energy versus pipeline baselines.
Bao \textit{et al.} \cite{bao2025dynamic} propose a dynamic routing framework for wireless edge-device LLM inference, combining a BERT-based semantic router with a latency model to balance quality and responsiveness. For multi-turn dialogues, it accounts for KV-cache recomputation overhead. Experiments show 5\%–15\% lower latency and 10\%–20\% fewer large-model calls, reducing communication and computation energy while maintaining accuracy. 

By dynamically routing tasks, adapting to thermal and resource conditions, and leveraging RL or quantization-aware strategies, these frameworks reduce redundant computation, communication overhead, strengthening Agentic AI through adaptive perception, efficient memory, faster reasoning, and proactive action.
\begin{table*}[tbp]
    \centering
    \caption{Summary of Computation Control Techniques (Section~\ref{sec:computation_control})}
    \label{tab:computation_control_summary}
    
     \scriptsize 
\setlength{\tabcolsep}{2pt} 
    
    \begin{tabularx}{\textwidth}{l|c|X|X|X}
        \hline
        \multicolumn{1}{c|}{\textbf{Technique}} & 
        \multicolumn{1}{c|}{\textbf{Ref.}} & 
        \multicolumn{1}{c|}{\textbf{Energy Saving Mechanism}} & 
        \multicolumn{1}{c|}{\textbf{Benefits}} & 
        \multicolumn{1}{c}{\textbf{Limitation}} \\ \hline

        \multirow{4}{*}{\shortstack[l]{\textbf{Token Length}\\ \textbf{Control}}} 
        & \cite{wei2025tokencommunicationeralarge} & Learns concise token representations via Generative Information Bottleneck (GenIB). & Reduces computational complexity and communication energy. & Minimal accuracy loss vs. efficiency gain. \\ \cline{2-5}
        
        & \cite{Zhang_2025} & Task-oriented multimodal token transmission with sliding-window pooling. & Balances transmission latency against model accuracy. &  Leads to $\sim$2.3\% performance degradation in low-SNR regimes. \\ \cline{2-5}
        
        & \cite{10778367} & Enforces optimal maximum token limits based on queuing theory. & Reduces queuing delay and user drop rates. & Risks truncating necessary reasoning steps (context loss). \\ \hline

        \multirow{5}{*}{\shortstack[l]{\textbf{Early Exit /}\\ \textbf{Adaptive Depth}}} 
        & \cite{11169709} & Partitions LLMs: high-confidence local exit, low-confidence offload to cloud. & Avoids unnecessary full-model inference and transmission. & Dependent on reliable confidence estimation. \\ \cline{2-5}
        
        
        &  \cite{zheng2025cloud} & Hierarchical early exit across end-edge-cloud. & Terminates inference once confidence is reached. & High scheduling complexity across heterogeneous devices.
 \\ \cline{2-5}
        
        & \cite{venkatesha2025fast}  & Speculative drafting on edge, verification on cloud with early exits. & 21\% speedup in vision-language control tasks. & Speedup relies on draft acceptance rate and network stability.
 \\ \hline

        \multirow{4}{*}{\shortstack[l]{\textbf{Layer}\\ \textbf{Skipping}}} 
        & \cite{10773726} & Distributed token-ring topology with skipping and recovery (JARVIS). & Enhances resilience to node failures and saves energy. & Depends on peer-level communication bandwidth and recovery overhead.
 \\ \cline{2-5}
        
        & \cite{11204255} & Input-conditioned encoder layer skipping (IDLD). & Outperforms random dropping; comparable to early exit. & Relies on the additional selection network's overhead and training.
 \\ \cline{2-5}
        
        & \cite{11166029} & Freezes lower layers and selectively trains upper layers. & Reduces communication costs by 70\% with $<$2\% accuracy loss. & May vary across specific clinical tasks and privacy settings.
 \\ \hline

        \multirow{6}{*}{\shortstack[l]{\textbf{Decoding}\\ \textbf{Simplification}}} 
        & \cite{10812936} & Speculative decoding with width-adaptive token trees. & Achieves up to 9.3$\times$ speedup without accuracy loss. & Be sensitive to the draft model's acceptance rate and task complexity.
 \\ \cline{2-5}
        
        
        & \cite{ning2025dssd} & Splits ``Accept/Reject'' (edge) and ``Resample'' (device). & 1.5-2.4$\times$ speedup vs. conventional distributed decoding. & Introduces additional downlink transmission overhead and device-side compute.
 \\ \cline{2-5}
        
        
        &  \cite{zhao2024edge} & Serial draft (terminal) + Parallel verification (edge). & Reduces delay and energy by 25\%-34\%. & Performance is sensitive to channel quality. \\ \hline

        \multirow{14}{*}{\shortstack[l]{\textbf{Workload}\\ \textbf{Scheduling}}} 
        & \cite{wilkins2024hybrid} & Hybrid framework allocating queries to CPU/GPU based on token length. & Reduces total CPU+GPU energy consumption by 7.5\%. & May not adapt to dynamic hardware states or shifting token distributions.
 \\ \cline{2-5}
        
        & \cite{stojkovic2025tapas} & Thermal-aware GPU VM placement (TAPAS). & Optimizes cooling and power constraints in datacenters. & Depends on the accuracy of historical power and temperature data.
 \\ \cline{2-5}
        
         & \cite{alizadeh2024duollm} & Dynamic token routing based on task complexity (Duo-LLM). & Reduces unnecessary energy use for simple tasks. & Produces suboptimal results compared to theoretical optima.
 \\ \cline{2-5}
        

        &  \cite{10571127} & Prioritizes requests with shorter output lengths. & Minimizes latency violations under energy budgets. & Depends on the accuracy of predicting varying user latency requirements. \\ \cline{2-5}
        
        & \cite{11095716} & Auction-based device selection and deadline-driven scheduling. & Reduces communication overhead by 54.7\%. &  Introduces additional negotiation latency and computational overhead for devices. \\ \cline{2-5}
        
        
        & \cite{li2024collm} & Tensor parallelism across devices (CoLLM). & 1.9-2.3$\times$ faster than hierarchical methods. & Optimal efficiency limited to ~4 devices. \\ \cline{2-5}
        
        
        
        & \cite{bao2025dynamic} & Semantic router + latency model for dynamic routing. & 5\%-15\% lower latency; 10\%-20\% fewer large-model calls. & Impacts the efficiency of dynamic routing decisions. \\ \hline
        

    \end{tabularx}
\end{table*}

\subsection{Input and Attention Optimization}  
\label{sec:input-opt}


Input and attention optimization reduces the overhead of long prompts and attention in LLMs, where cost scales quadratically with sequence length~\cite{keith2024dtp}. Strategies include pruning redundant tokens, sparsifying attention, compressing KV caches, and reusing states to avoid unnecessary operations.  
For Agentic AI inference, efficiency improves through token pruning~\cite{keith2024dtp}, sparse attention for long sequences~\cite{singhania2024loki}, and KV caching/reuse to cut recomputation and memory~\cite{liu2024cachegen}. These methods lower computation, communication, and storage, enabling faster, more energy-efficient, and sustainable deployment in constrained networks.

\subsubsection{\textbf{Token pruning}}  

Token pruning reduces the number of tokens processed during inference by selectively discarding those deemed less important. In Agentic AI systems, long prompts and multimodal inputs often contain redundant or low-value tokens that increase computational workload, memory access, and communication overhead. By pruning such tokens, models can lower the number of matrix multiplications, reduce DRAM transfers, and minimize wireless transmission costs in distributed networks~\cite{jiang2023llmlingua}. This directly translates into reduced energy consumption, faster inference, and improved feasibility of Agentic AI deployment.  

Token pruning frameworks across text, vision, and multimodal LLMs demonstrate their effectiveness in reducing overhead and enabling efficient inference.
Jiang \textit{et al.}~\cite{jiang2023llmlingua} propose LLMLingua, a coarse-to-fine compression framework that prunes redundant input tokens. In mobile edge scenarios with cloud-offloading, LLMLingua reduces prompt size by up to 20$\times$, lowering wireless transmission latency and prefill computational energy, while maintaining semantic integrity.  
Wang \textit{et al.}~\cite{wang2021spatten} present SpAtten, which progressively discards ``lazy'' tokens based on cumulative attention scores. This reduces expensive matrix multiplications and DRAM accesses, achieving orders-of-magnitude improvements in energy efficiency on mobile hardware compared to unpruned inference. 
Zhong \textit{et al.}~\cite{zhong2025aim} propose AIM, a training-free adaptive inference method for multi-modal LLMs, which merges redundant visual tokens before the LLM and prunes less important ones inside layers using PageRank on attention weights. On video benchmarks, AIM reduces FLOPs from 99.63~TB to 14.76~TB ($\approx 6.8\times$) and prefill time from 439.6~ms to 55.0~ms ($\approx 8\times$) while retaining 99.7\% of the baseline accuracy.


By eliminating redundant tokens, these token pruning methods reduces computation, memory, and communication costs, enabling Agentic AI with sharper perception, faster reasoning, and proactive action in edge and mobile networks.

\subsubsection{\textbf{Sparse attention}}  

Sparse attention reduces the quadratic complexity of standard self-attention by restricting computations to a subset of tokens or by leveraging low-rank approximations~\cite{zhou2021informer}. In Agentic AI inference, especially for long-context or multimodal tasks, dense attention leads to high computational cost, memory traffic, and energy consumption. Sparse attention mitigates these issues by lowering the number of matrix multiplications, reducing DRAM access, and minimizing I/O overhead, thereby enabling faster and more energy-efficient inference. 

Recent studies illustrate different strategies and benefits of sparse attention. 
Dao \textit{et al.}~\cite{dao2022flashattention}  introduce FlashAttention, an I/O-aware exact
attention algorithm that accelerates Transformer training and
inference by minimizing costly GPU high-bandwidth memory
(HBM) accesses. Using tiling and recomputation, FlashAttention avoids materializing the full attention matrix; instead
it leverages fast on-chip static random-access memory (SRAM) and fused CUDA kernels to
reduce memory traffic.
Hadish \textit{et al.}~\cite{11059615} integrate FlashAttention and ProbAttention into a dual-path Transformer for microgrid scenarios. This design achieves efficient parallel computing and low memory complexity, surpassing baselines under resource constraints.  
Tanwar \textit{et al.}~\cite{tanwar2025energy} propose M7BCO, which combines sparse autoencoders with Grouped Query Attention and RMSNorm. By penalizing redundant transmissions and optimizing routing, it reduces energy consumption in wireless sensor networks while enabling adaptive decision-making: The proposed method achieves 20.2 mJ, 61.5 mJ, and 99.4 mJ energy consumption for 150, 300, and 500 nodes, respectively, outperforming CORP (22.5, 65.3, 100.2 mJ).

By reducing computational complexity and memory traffic, sparse attention enables energy‑efficient, high‑performance Agentic AI across diverse edge networks.  

\subsubsection{\textbf{KV caching and reuse}}  

KV caching and reuse optimize attention in LLMs by reducing memory and compute overhead. During inference, self-attention repeatedly accesses KV pairs, quickly exhausting GPU/DRAM and raising energy use from transfers and recomputation~\cite{li2025hotprefix}. Compressing, reusing, or selectively loading KV caches cuts memory footprint, I/O, and redundant computation~\cite{liu2025tinyserve}. In Agentic AI inference, especially at the edge, such techniques lower energy demand, accelerate responses, and support longer contexts under resource constraints.

For instance, Luo \textit{et al.}~\cite{luo2025simllm} propose Sim-LLM, which reuses KV caches across semantically similar tasks using cosine similarity and locality-sensitive hashing (LSH) mapping. This reduces memory consumption and accelerates inference in both single-node and multi-node edge deployments.  
Zhang \textit{et al.}~\cite{zhang2023h2o} propose H2O, which retains only ``Heavy Hitter'' tokens in memory. This reduces the footprint by up to 5$\times$, enabling longer sequence generation on memory-constrained mobile hardware without out-of-memory errors.  
Liu \textit{et al.}~\cite{liu2024kivi} propose KIVI, which applies asymmetric quantization (2-bit for values, 4-bit for keys). This reduces KV memory by 2.6$\times$, enabling mobile devices to support up to 64K tokens without crashing.  
Tang \textit{et al.}~\cite{tang2024quest} introduce Quest, which loads only critical KV cache pages based on query vectors. This reduces memory movement and achieves over 2$\times$ speedup in self-attention, lowering energy costs for long-context inference.  

By reducing memory footprint, minimizing I/O overhead, and avoiding redundant computation, these methods enable long-context and multi-turn tasks to run on resource-constrained edge devices, while enhancing Agentic AI capabilities, with an emphasis on efficient memory utilization.

\begin{table*}[tbp]
    \centering
    \caption{Summary of Input and Attention Optimization Techniques (Section~\ref{sec:input-opt})}
    \label{tab:input_attention_optimization}
    
    \scriptsize
    \renewcommand{\arraystretch}{1} 
    
    \begin{tabularx}{\textwidth}{l|c|X|X|X}
        \hline
        \multicolumn{1}{c|}{\textbf{Technique}} & 
        \multicolumn{1}{c|}{\textbf{Ref.}} & 
        \multicolumn{1}{c|}{\textbf{Energy Saving Mechanism}} & 
        \multicolumn{1}{c|}{\textbf{Benefits}} & 
        \multicolumn{1}{c}{\textbf{Limitation}} \\ \hline

        \multirow{6}{*}{\shortstack[l]{\textbf{Token}\\ \textbf{Pruning}}} 
        & \cite{jiang2023llmlingua} & Coarse-to-fine compression pruning redundant tokens (LLMLingua). & Reduces prompt size up to 20$\times$; lowers transmission/prefill energy. & Potential loss of fine-grained context details. \\ \cline{2-5}
        
        & \cite{wang2021spatten} & Progressively discards ``lazy'' tokens based on cumulative attention. & Orders-of-magnitude energy efficiency improvement on mobile. & Cause loss of subtle logical nuances in complex reasoning tasks. \\ \cline{2-5}
        
        
        & \cite{zhong2025aim} & Merges redundant visual tokens and prunes via PageRank (AIM). & Reduces FLOPs (6.8$\times$) and prefill time (8$\times$) with 99.7\% accuracy. & May over-simplify visual features in texture-heavy scenes. \\ \hline

        \multirow{6}{*}{\shortstack[l]{\textbf{Sparse}\\ \textbf{Attention}}} 
        & \cite{dao2022flashattention} & I/O-aware tiling and recomputation to minimize HBM access (FlashAttention). & Accelerates inference; reduces memory traffic via SRAM usage. & Dependent on GPU memory hierarchy and SRAM size. \\ \cline{2-5}
        
        
        & \cite{11059615} & Integrates FlashAttention and ProbAttention for smart grids. & Low memory complexity suitable for constrained resources. & Struggles with long-term seasonal shifts. \\ \cline{2-5}
        
        & \cite{tanwar2025energy} & Sparse autoencoders with Grouped Query Attention (M7BCO). & Reduces energy consumption in WSNs (20.2 mJ vs 22.5 mJ). & Latency may exceed the real-time demands of fast-fading channels.
 \\ \hline

        \multirow{10}{*}{\shortstack[l]{\textbf{KV Caching}\\ \textbf{\& Reuse}}} 
        & \cite{luo2025simllm} & Reuses KV caches across semantically similar tasks via LSH (Sim-LLM). & Accelerates inference in multi-node edge deployments. & KV cache reuse is contingent on high semantic similarity between tasks. \\ \cline{2-5}
        
        
        & \cite{zhang2023h2o} & Retains only ``Heavy Hitter'' tokens in memory (H2O). & Reduces footprint by 5$\times$; enables longer sequence generation. & Static ratios fail on unconventional text distributions.
 \\ \cline{2-5}
        
        & \cite{liu2024kivi} & Asymmetric quantization: 2-bit value, 4-bit key (KIVI). & Reduces memory by 2.6$\times$; supports 64K tokens on mobile. & May lose precision in complex long-context reasoning.
 \\ \cline{2-5}
        
        & \cite{tang2024quest} & Loads only critical KV cache pages based on query vectors (Quest). & $>2\times$ speedup in self-attention; lowers I/O energy. &  Accuracy relies on effective Top-K critical page selection.\\ \hline
        
    \end{tabularx}
\end{table*}
\subsection{Hardware-Aware Inference}
\label{sec:hardware-inf}

Hardware-aware inference optimizes LLMs by accounting for hardware limits, e.g., CPUs, GPUs, neural processing units (NPUs), field-programmable gate arrays (FPGAs) and memory. Unlike algorithm-only methods, it co-designs inference with hardware scheduling, precision control, and memory management to maximize throughput and minimize energy~\cite{kakolyris2024sloaware,kwon2023pagedattention}.  
For Agentic AI, energy depends not only on FLOPs but also on hardware utilization. Poor mapping wastes compute, increases transfers, and causes thermal throttling. Tailoring inference yields better energy–performance trade-offs via precision scheduling~\cite{frantar2025marlin}, DVFS~\cite{kakolyris2025throttll}, and memory/I/O optimization~\cite{jiang2025kvpr}. Precision scheduling assigns mixed‑precision formats, e.g., FP16, INT8, or BF16, across layers to cut memory and accelerate operations. DVFS adjusts voltage/frequency to balance energy and latency. Memory/I/O optimization reduces costly CPU–GPU transfers and storage reads. These strategies enable scalable, energy-efficient Agentic AI deployment.

\subsubsection{\textbf{Precision scheduling}} 
Precision scheduling refers to the dynamic selection and allocation of numerical precision (e.g., FP8, FP16, BF16, and INT4) across layers, tasks, or devices during inference~\cite{frantar2025marlin}. Because different components of LLMs vary in their sensitivity to quantization, this approach enables fine‑grained control over computational accuracy and energy efficiency. In Agentic AI inference, precision scheduling is critical: Lower precision reduces memory footprint, accelerates matrix multiplications, and minimizes energy consumption, while higher precision can be reserved for critical layers to ensure reliability. By adaptively balancing precision levels, systems achieve sustainable performance under strict energy and latency constraints, especially in edge and heterogeneous networks.  

Recent studies demonstrate how adaptive precision scheduling across attention, decoder blocks, and heterogeneous accelerators can drastically reduce memory, latency and energy, while preserving accuracy for LLM deployment.
Li \textit{et al.}~\cite{li2023llmmq} design LLM-MQ with sensitivity‑based precision allocation. Instead of uniformly quantizing all layers to the same low bit‑width, LLM-MQ measures each layer’s sensitivity to quantization error using gradient information and allocates higher precision (e.g., 4‑bit) to sensitive layers while assigning lower precision (e.g., 2‑bit) to less sensitive ones, under a global memory budget formulated as an integer programming problem, achieving fast LLM inference on memory and energy-constrained edge devices. 
Bajpai \textit{et al.}~\cite{bajpai2025ecollm} propose EcoLLM for edge systems, integrating mixed precision with structured pruning. Using an analytical E‑Model to predict energy and an E‑Metric to balance efficiency and accuracy, EcoLLM adaptively assigns bit‑widths and sparsity across layers: Critical layers retain higher precision (e.g., INT8) and low sparsity, while less sensitive layers use aggressive compression (e.g., INT2 with high sparsity). This non‑uniform scheduling achieves up to 6.4$\times$ compression and 45\% power reduction on GPT‑2 with minimal accuracy loss.
Cho \textit{et al.}~\cite{11218199} assign bit‑widths via Hessian‑ and norm‑based sensitivity. Outlier columns stay FP16, others quantized to INT2/3/4. Scaled power‑of‑two logarithmic quantization improves resolution near zero and hardware efficiency. A dual‑mode matrix multiplication unit switches between FP16 and logarithmic datapaths, achieving 11.8$\times$ compression and 1.82$\times$ energy efficiency with accuracy close to full precision.

By adaptively balancing precision across layers, tasks, and devices, these methods reduce energy consumption, memory usage, and latency while preserving accuracy. This strengthens Agentic AI capabilities, most notably efficient memory utilization, which supports reliable context retention and enables downstream reasoning and action in heterogeneous networks.

\subsubsection{\textbf{DVFS}}  


DVFS is a hardware-level energy management technique that adjusts processor voltage and frequency at runtime to balance performance and energy~\cite{ye2025agft}. In Agentic AI inference, workloads vary across prompt processing, token generation, and multimodal tasks. Prefill is compute-intensive, while decode is dominated by KV-cache lookups. Without DVFS, processors run at fixed high frequencies, wasting energy and causing thermal stress. By tuning voltage and frequency to workload intensity and CPU–GPU coupling, DVFS reduces power draw, mitigates overheating, and improves efficiency while maintaining latency~\cite{ye2025agft}.

Recent studies highlight the wide-ranging applications and benefits of DVFS in LLM inference across mobile devices, edge networks, and large-scale datacenters.
Zhang \textit{et al.}~\cite{zhang2025dissecting} analyze interactions among CPU, GPU, and memory DVFS regulators in mobile devices, revealing a ``downward spiral'' effect: When the GPU waits for CPU kernels, utilization drops, prompting both governors to lower frequencies, cascading into higher latency and energy inefficiency. To address this, they propose FUSE, a unified governor jointly managing CPU, GPU, and memory. Offline profiling identifies optimal triplets of frequencies that minimize latency under fixed energy or vice versa. For example, in TinyLlama-1.1B’s decode stage, the default GPU governor at 424\,MHz yields 215.1\,ms/token at 402.7\,mJ, while pinning at 848\,MHz reduces latency to 126.9\,ms (41\% faster) with similar energy (396.5\,mJ). Exploiting LLM inference characteristics (compute‑intensive prefill, batch‑size‑1 decode, and tight CPU–GPU coupling), FUSE fixes components at optimal frequencies at runtime, reducing latency by 7\%–36.8\% under the same energy budget.
Patel \textit{et al.}~\cite{patel2024polca} characterize LLM inference power use in cloud datacenters, noting prompt processing is compute‑intensive while token sampling is memory‑bound. Unlike training clusters that sustain peak GPU power, inference clusters rarely hit maximum draw, leaving capacity for safe oversubscription. Building on this, they propose POLCA, which overlaps GPU frequency locking and power capping with inference workloads. Tailored to LLM phases, POLCA enables 30\% more servers under the same power budget with minimal performance loss, showing how workload‑aware design alleviates GPU constraints.
Kurma \textit{et al.}~\cite{11152831} integrate DVFS into a 6G intelligent medical network framework. By dynamically adjusting CPU frequency according to task
intensity at IoT devices, they balance local energy consumption with execution time, reducing cumulative overhead by 44.96\% compared to manual optimization. This demonstrates DVFS's role in energy-efficient edge computing for critical applications. 


By adaptively tuning processor voltage and frequency, DVFS reduces energy consumption, alleviates thermal bottlenecks, and improves throughput across mobile, edge, and cloud environments, thereby enhancing Agentic AI reasoning and action under dynamic resource conditions.

\subsubsection{\textbf{Memory and I/O optimization}}  

Memory and I/O optimization focuses on reducing the overhead of data movement between CPU, GPU, and external storage, as well as improving memory utilization during inference~\cite{jiang2025kvpr}. The optimization in Agentic AI is driven by their unique inference characteristics, such as large parameter sizes, autoregressive decoding, and KV-cache management. Unlike generic workloads, inference in Agentic AI repeatedly accesses massive KV caches and loads billions of weights across transformer blocks, making data movement often more expensive than computation itself~\cite{jiang2025kvpr}. Recent studies therefore design LLM-specific optimizations. 

Zhao \textit{et al.}~\cite{zhao2024hetegen} propose HeteGen, a CPU--GPU heterogeneous reasoning system for LLMs. Since linear layers dominate memory (e.g., OPT-30B, 97\%), and autoregressive decoding requires small batches, HeteGen overlaps CPU computation with GPU communication and applies hybrid parallelism. Parameters reside in CPU memory while critical weights stream to GPU, reducing transfers and idle time. Aligning strategies with prefill and decode stages, HeteGen achieves up to 317\% speedup on constrained devices and enables large models within limited GPU memory.  
Alizadeh \textit{et al.}~\cite{alizadeh2024llm} present LLM in a Flash, a memory-aware system exploiting activation sparsity ($>$90\% in Feed-Forward Network (FFN) layers). Parameters are stored in flash and only active subsets loaded into DRAM. Techniques like \emph{windowing} (reusing recent neurons) and \emph{row--column bundling} align flash reads with FFN access patterns. Tailored I/O scheduling enables models twice DRAM size to run efficiently, achieving up to 4$\times$ CPU and 20$\times$ GPU speedups while reducing redundant transfers.  
Li \textit{et al.}~\cite{11119787} propose TPI-LLM, a tensor-parallel inference framework for large transformers on edge devices. Unlike pipeline parallelism, it distributes attention heads and FFN weights across devices. A sliding-window scheduler overlaps disk I/O with computation, exploiting autoregressive decoding where only subsets of weights and KV-cache are needed. To mitigate link latency, TPI-LLM uses a star-based all-reduce optimized for small-token exchanges, reducing memory footprint and energy use for multi-device serving.  

These techniques exploit the distinction between prefill (compute‑intensive, weight‑heavy) and decode (latency‑sensitive, cache‑heavy), tailoring memory scheduling to LLM workloads. By reducing redundant KV‑cache transfers, streaming only required weights, and embedding compute into memory, they lower energy use and enable efficient long‑context inference on constrained hardware. The key enhancement lies in Agentic AI’s memory function, which improves context retention and state management while supporting perception, reasoning, and action through proactive resource use.
\begin{table*}[tbp]
    \centering
    \caption{Summary of Hardware-Aware Inference Techniques (Section~\ref{sec:hardware-inf})}
    \label{tab:hardware_aware_inference}
    
    \scriptsize
    \renewcommand{\arraystretch}{1} 
    
    \begin{tabularx}{\textwidth}{l|c|X|X|X}
        \hline
        \multicolumn{1}{c|}{\textbf{Technique}} & 
        \multicolumn{1}{c|}{\textbf{Ref.}} & 
        \multicolumn{1}{c|}{\textbf{Energy Saving Mechanism}} & 
        \multicolumn{1}{c|}{\textbf{Benefits}} & 
        \multicolumn{1}{c}{\textbf{Limitation}} \\ \hline

        \multirow{6}{*}{\shortstack[l]{\textbf{Precision}\\ \textbf{Scheduling}}} 
        
        & \cite{li2023llmmq} & Allocates precision based on layer sensitivity. & Ensures accuracy under memory budgets of edge devices. &  Increases scheduling and workload balancing complexity. \\ \cline{2-5}
        
        & \cite{bajpai2025ecollm} & Integrates mixed precision with structured pruning (EcoLLM). & Avoids reconstruction errors; efficient compression for edge. & Increase implementation and hardware mapping complexity.
 \\ \cline{2-5}


        & \cite{11218199} & Sub-4-bit inference with Dual-Mode Matrix Multiplication Unit. & 1.82$\times$ higher energy efficiency; 11.8$\times$ compression. & Performance relies on specific log-scale structured sparsity. \\ \hline

        \multirow{5}{*}{\textbf{DVFS}} 
        & \cite{zhang2025dissecting} & Unified CPU-GPU-Memory governor identifying optimal freq triplets (FUSE). & Reduces latency by 7\%-36.8\% under same energy budget. & Requires specialized hardware to achieve the claimed energy efficiency. \\ \cline{2-5}
        
        & \cite{patel2024polca} & Overlaps GPU power capping with inference phases (POLCA). & Deploys 30\% more servers under same power budget. & Limits portability across different mobile SoC models. \\ \cline{2-5}

        & \cite{11152831} & IAI-LLM driven CPU frequency scaling and RIS beamforming. & Minimizes cumulative overhead by 44.96\% via task-aware DVFS. & Frequent frequency switching increases hardware signaling overhead. \\ \cline{1-5}

        \multirow{8}{*}{\shortstack[l]{\textbf{Memory}\\ \textbf{\& I/O Opt.}}} 
        & \cite{zhao2024hetegen} & Overlaps CPU computation with GPU communication (HeteGen). & 317\% speedup; runs large models on constrained devices. & High training costs and complexity in large-scale IoT networks. \\ \cline{2-5}
        
        & \cite{alizadeh2024llm} & Stores params in flash, loads active subsets to DRAM (LLM in a Flash). & Runs models 2$\times$ larger than DRAM capacity; 20$\times$ GPU speedup. & Flash I/O bandwidth is the bottleneck. \\ \cline{2-5}
        
        & \cite{11119787} & Sliding-window memory scheduler overlapping disk I/O (TPI-LLM). & Enables 70B-scale LLM serving on low-resource edge. & Accelerate hardware wear and increase latency.\\ \hline 
        

    \end{tabularx}
\end{table*}

\subsection{Lessons Learned}
From the surveyed optimization methods, several key lessons emerge. First, no single technique suffices. Energy efficiency requires a multi-layered approach that integrates compression, adaptive computation, and hardware co-design. Second, trade-offs are inevitable: Aggressive quantization or pruning may yield substantial energy savings but can degrade reasoning accuracy, highlighting the need for task-aware and adaptive strategies \cite{ieee2024pruning}. Third, networking-aware design is essential. Simplified models not only reduce computation but also lower communication overhead, which is critical in federated and multi-agent settings \cite{xie2025edge}. Finally, the most promising direction lies in cross-layer co-optimization, where algorithmic, architectural, and networking strategies are jointly tuned to achieve sustainable Agentic AI inference \cite{liu2024energymode}. These lessons underscore that energy-efficient optimization is not merely a technical add-on but a foundational requirement for scalable, autonomous intelligence in real-world environments.

\section{Integrated Wireless-Edge Intelligence for Sustainable Agentic AI}
\label{sec:crosslayer} 
The deployment of Agentic AI in wireless and edge networks introduces unique challenges, as both computation and communication contribute significantly to overall energy consumption. Joint optimization across AI inference, wireless transmission, and edge resource allocation has emerged as a critical research direction. This section reviews key approaches and design principles, categorizing them into three synergistic themes: cross-layer optimization of interdependent variables, collaborative edge–cloud execution, and integrated communication–inference co-design.

\begin{figure}[t]
    \centering
\includegraphics[width=0.7\textwidth]{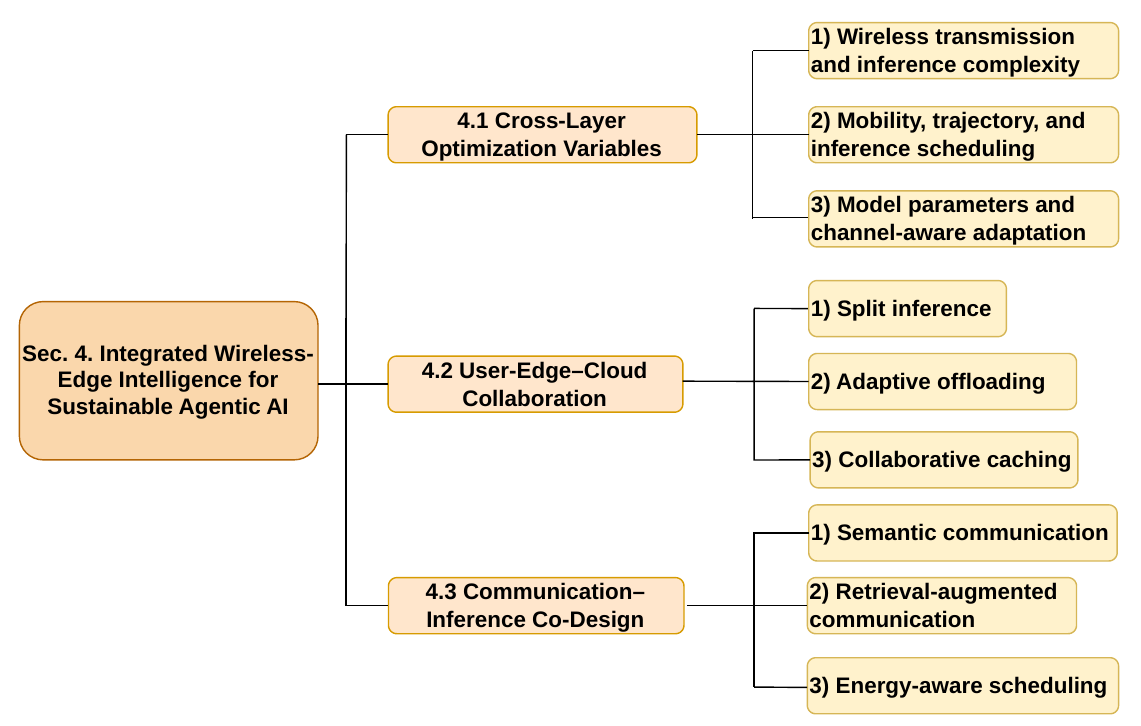} 
    \caption{An overview of integrated wireless-edge intelligence for sustainable Agentic AI.}
    \label{fig:sec4}
\end{figure}

\subsection{Cross-Layer Optimization Variables}
Achieving energy efficiency in distributed Agentic AI requires moving beyond isolated model or communication optimizations. Cross‑layer optimization couples inference with the wireless stack, co‑adapting variables in real time. Joint tuning balances computation, communication, and accuracy, addressing energy bottlenecks in mobile and edge deployments. The principle is to avoid over‑provisioning by dynamically aligning compute and communication states. Key jointly tunable variables are outlined below.
\subsubsection{ \textbf{Wireless transmission and inference complexity}}
This co-optimization axis links transmission parameters (power, MCS, bandwidth) to energy, latency, and reliability in wireless inference. Settings must adapt to task demands: Long sequences require high bandwidth and robust modulation, while shorter or compressed features lower communication cost. The strategy is to align communication with inference complexity under dynamic channels, balancing compute and transmission energy.
He \textit{et al.}~\cite{he2024active} propose an active inference framework for LLM offloading in cloud-edge systems. Instead of reward-driven DRL, it leverages the free energy principle for data-efficient policy learning across bandwidth, compute, and memory. By partitioning tasks adaptively, the framework reduces latency and energy, achieving faster convergence and better accuracy-latency trade-offs with models like GPT-J-6B compared to mainstream DRL

\subsubsection{\textbf{Mobility, trajectory, and inference scheduling}}
For mobile AI agents such as UAVs, robots, or connected vehicles, physical movement is a key optimization lever~\cite{zhang2019joint,wang2020joint}. The agent's trajectory determines its communication distance from access points or other agents, affecting path loss, signal strength, and thus the transmission energy required for a given data rate. Poorly planned routes can trap agents in weak coverage, forcing high transmit power or buffering, which increases latency and energy~\cite{zeng2016uav}. Mobility also shapes the spatial and temporal distribution of tasks. An agent can schedule data collection, perception, and inference tasks to coincide with locations where communication is favorable or where edge compute resources are physically proximate. By co-optimizing the physical flight path/route with the timing and location of computation-intensive inference tasks, the system can minimize the product of communication energy (by reducing distance and avoiding obstacles) and redundant computation (by planning sensing actions to be efficient).

\subsubsection{\textbf{Model parameters and channel-aware adaptation}}
AI model parameters can adapt to networking conditions. Unlike static compression, cross‑layer optimization enables \textit{runtime} tuning of pruning, quantization, and sparsity (e.g., MoE activation). Model complexity becomes a knob tied to energy and link quality. Strong channels allow larger, precise variants for accuracy, while limited bandwidth or battery trigger compressed versions. This reduces computation and transmission, saving energy with context‑dependent accuracy trade‑offs.
Wang \textit{et al.}~\cite{zhou2024decentralized} consider networking conditions when leveraging MoE models in edge networks. Their dynamic gating mechanism selects experts not only by input data but also by available bandwidth, activating communication-friendly experts under congestion to mitigate stragglers and balance accuracy with latency. 
Bai \textit{et al.}~\cite{bai2025adaptive} propose a weight-shared dynamic network supporting multi-dimensional compression (pruning, downsampling, quantization) at partition points in a co-inference pipeline. A dynamic programming scheduler jointly determines the partition point, compression parameters, and resource allocation, optimizing intermediate features for bandwidth and latency constraints to improve energy efficiency. 

In summary, cross-layer optimization provides actionable handles across computation and communication. By jointly tuning transmit mode with inference length, scheduling with trajectory, and model sparsity with channel state, Agentic AI achieves holistic energy efficiency beyond isolated methods. This requires feedback loops and controllers aware of both task graph and network state, enabling sustainable autonomy under dynamic, resource-constrained conditions.

\subsection{User-Edge–Cloud Collaboration}
Joint optimization in Agentic AI extends beyond offloading, evolving into three‑tier collaboration among user devices, edge servers, and the cloud. This hierarchical orchestration flexibly balances low latency, high accuracy, and strict energy efficiency. By strategically deciding computation placement, the system mitigates resource limits while leveraging scalable but energy‑intensive cloud infrastructure. Collaboration unfolds through three strategies: split inference, adaptive offloading, and collaborative caching. 

\subsubsection{\textbf{Split inference}} 
Split Inference distributes execution of large AI models across device, edge, and cloud. Lightweight layers (e.g., feature extraction) run on the device, minimizing local computation, battery drain, and raw data transfer. Intermediate layers are processed at the edge, leveraging nearby resources to reduce latency and avoid long-distance costs. The deepest, most resource-intensive layers execute in the cloud, where GPU/TPU clusters deliver high efficiency. Crucially, the partition point adapts dynamically to bandwidth, latency, and battery, balancing energy savings with performance.
Chen \textit{et al.}~\cite{chen2025adaptive} propose an adaptive layer-splitting framework for wireless LLM inference. Using Model-Based Reinforcement Learning (MBRL), it dynamically identifies the best split point between user and edge, with a reward surrogate model balancing accuracy and energy under volatile channels. Similarly, Ma \textit{et al.}~\cite{ma2025multi} present MMSL, a two-stage scheduling framework for collaborative LLM inference across multi-tier cloud-edge networks.  Integer Linear Programming (ILP) optimizes inter-tier partitioning, while GNNs allocate tasks intra-tier. By reassigning layers based on real-time node conditions, MMSL minimizes latency and improves energy utilization across heterogeneous resources.

\subsubsection{\textbf{Adaptive offloading}} Adaptive offloading makes fine-grained, real-time decisions on whether tasks or operators run locally or remotely. Unlike split inference at the layer level, it operates on individual operators or tokens, offering greater flexibility to optimize energy while maintaining responsiveness. The decision is based on a multi-factor analysis, including real-time CSI, device battery level, and the urgency or complexity of the task.
Xue \textit{et al.}~\cite{xue2025wdmoe} propose Wireless Distributed MoE (WDMoE) for LLMs, placing gating and attention at edge servers while distributing experts to devices, jointly optimizing expert selection and bandwidth with a latency-aware metric. 
Hao \textit{et al.}~\cite{hao2024hybrid} introduce token-level collaboration: Small models generate most tokens locally, while the cloud verifies ``hard'' tokens. This scheme achieves near-LLM quality at only 25\%--31\% of the cost, demonstrating the potential of fine-grained collaboration for energy-efficient inference.


\subsubsection{\textbf{Collaborative caching}} Collaborative caching aims to reduce redundant computation and communication by storing and reusing intermediate results, feature embeddings, or retrieved knowledge at the user devices or network edge. When multiple agents or users request similar inferences, cached results can be served directly, bypassing the need to re-execute the full model or retransmit large amounts of data. This mechanism not only lowers latency but also significantly cuts energy consumption by avoiding duplicate workloads and reducing unnecessary cloud interactions.
Beyond simple reuse, collaborative caching introduces semantic awareness and adaptive placement. 
Pour \textit{et al.}~\cite{pour2025meancache} propose a semantic-aware caching framework leveraging federated learning to identify and store semantically equivalent queries locally. By recognizing query similarity across users, the system bypasses redundant LLM invocations, saving computational cycles and transmission energy. 
{\color{black}Liu \textit{et al.}~\cite{liu2025joint} study joint model caching and resource allocation for Generative AI in wireless edge networks. Their DDPG-based algorithm dynamically optimizes model placement, bandwidth, and computational resources (e.g., denoising steps) to balance service latency and content quality.}

User-edge-cloud collaboration transforms classical offloading into an energy-aware partnership. Split inference adapts the model graph to network conditions, while adaptive offloading makes granular task-placement decisions. Collaborative caching leverages collective intelligence to avoid redundant work. These strategies enable sustainable, energy-efficient deployment of Agentic AI across the continuum from device to edge to cloud.

\subsection{Communication–Inference Co-Design}  
Agentic AI relies on frequent agent--server interactions, making communication overhead critical. Sustainable performance under strict latency and energy limits requires co-design of communication and inference. By coupling transmission with execution, systems cut redundant workloads, minimize bandwidth, and improve energy use while preserving accuracy. Three representative directions are semantic communication, retrieval‑augmented communication, and energy‑aware scheduling.

\subsubsection{\textbf{Semantic communication}}  
In Agentic AI systems, semantic communication shifts from transmitting raw data to exchanging compressed, meaning-centric representations. Instead of sending entire signals, the transmitter encodes essential semantics (e.g., prompts, embeddings, high-level features), while the receiver’s generative capabilities reconstruct high-fidelity content. This reduces bandwidth and transmission energy, transforming communication into a knowledge-centric process that directly supports inference tasks.  
For example, Liang \textit{et al.}~\cite{liang2025generative} present a generative semantic communication framework where raw data is encoded into compact prompts before transmission, and receivers reconstruct outputs via generative models. This reduces bandwidth, latency, and transmission energy by avoiding redundant exchange.  
Similarly, Guo \textit{et al.}~\cite{guo2023semantic} propose an importance‑aware scheme that quantifies token contributions with pre-trained
language models. By transmitting only high‑score tokens and filtering redundancy, their method compresses bandwidth, lowers downstream workload, and conserves energy.

\subsubsection{\textbf{Retrieval-augmented communication}} 
Retrieval-augmented communication extends knowledge-centric transmission by letting agents exchange retrieved artifacts instead of full queries or raw contexts. Rather than sending large data volumes, agents share intermediate results such as KV caches, embeddings, and retrieved passages, which can be reused collaboratively to avoid repeated inference and minimize communication energy. Chen \textit{et al.}~\cite{chen2025kvcomm} propose KVCOMM, enabling agents to transmit and reuse KV caches of overlapping contexts. By bypassing redundant prefill computations, KVCOMM accelerates collaborative inference and reduces energy otherwise consumed by repeated token processing. Tang \textit{et al.}~\cite{tang2025retrieval} introduce a retrieval-augmented semantic communication framework integrating RAG into generative AI. Instead of transmitting entire datasets, the receiver reconstructs information by retrieving relevant contexts from external knowledge bases, reducing transmission overhead and avoiding duplication, thereby conserving both communication and inference energy.
\subsubsection{\textbf{Energy-aware scheduling}}  
Efficient communication--inference co-design requires dynamic scheduling that accounts for energy budgets, idle cycles, and heterogeneous resources. Rather than treating requests as a static pipeline, energy-aware scheduling coordinates inference in real time, balancing network load, exploiting batching, and adapting to device capabilities. This keeps services responsive while aligning operation with sustainability goals.  
Zhang \textit{et al.}~\cite{zhang2025beyond} show how batching and joint resource allocation reduce per-request energy by maximizing utilization and minimizing redundancy. Rajashekar \textit{et al.}~\cite{rajashekar2025toward} study carbon-aware routing for LLM inference that distributes requests by real-time energy profiles. By batching idle GPU cycles, their approach balances load while minimizing carbon footprint and per-token energy, showing how intelligent scheduling reduces latency and energy.

Communication--inference co-design turns communication from passive pipeline into active optimization. Semantic communication reduces overhead with compressed representations. Retrieval-augmented communication saves energy by reusing artifacts. Energy-aware scheduling balances load and aligns inference with sustainability goals. These strategies enable Agentic AI to deliver low latency, high accuracy, and energy-efficient performance under strict resource constraints.

\subsection{Lessons Learned}

The exploration of joint optimization strategies across wireless networks, edge computing, and AI inference reveals several insights that can guide future research and system design for energy-efficient Agentic AI deployment. First, synergy dominates isolation. Jointly tuning wireless transmission, model complexity, and scheduling decisions creates energy efficiency unattainable through isolated optimizations~\cite{he2024active, zhou2024decentralized}. Second, adaptability is essential. The optimal granularity of collaboration, whether layer-level splitting~\cite{chen2025adaptive}, token-level offloading~\cite{yang2024efficient}, or result-level caching~\cite{pour2025meancache}, depends on channel conditions, device capabilities, and task urgency. Third, semantic efficiency transcends bit efficiency. Exchanging compact prompts, KV caches~\cite{chen2025kvcomm}, or retrieved knowledge~\cite{tang2025retrieval} rather than raw data reduces both communication and downstream computation energy. Finally, robustness must be co-designed. Wireless volatility and agent mobility can rapidly obsolete offloading decisions, necessitating predictive channel models and seamless state migration mechanisms~\cite{zhang2019joint}. These lessons underscore that sustainable networked Agentic AI requires not merely stacking optimizations, but reconceptualizing the interplay between intelligence and infrastructure.

\section{Open Challenges and Future Directions}
\label{sec:future}

Despite advances in energy-efficient Agentic AI inference, several challenges remain unresolved. These challenges span the tension between reasoning quality and resource constraints, the complexity of multi-agent coordination, and the need for sustainable deployment at scale. Addressing these gaps requires novel approaches that reimagine the relationship between intelligence, energy, and infrastructure.

\subsection{Fundamental Trade-offs}
\begin{itemize}
[leftmargin=*]
    \item 
\textbf{Balancing reasoning reliability with energy efficiency.} A persistent tension exists between the depth of reasoning required for reliable autonomous decision-making and the energy costs of iterative inference. Current early exit and adaptive depth mechanisms (Section~\ref{sec:computation_control}) often rely on heuristic confidence thresholds that may fail in high-stakes scenarios. Recent work~\cite{jazbec2024early} highlights that standard uncertainty quantification techniques like Bayesian methods or conformal prediction can lead to inconsistency across exits in early-exit neural networks. Future work must develop uncertainty-quantified adaptive reasoning frameworks that explicitly model the risk of premature termination against energy savings, potentially leveraging anytime-valid confidence sequences (AVCSs) to maintain consistency across exits while providing statistical guarantees on reasoning quality under energy budgets.

\item \textbf{Enabling cross-modal and cross-agent collaboration.} Future Agentic AI systems will increasingly integrate heterogeneous modalities (vision, language, audio, sensor data) and coordinate across distributed agents. However, cross-modal fusion incurs substantial energy overhead from aligning disparate representations, while multi-agent collaboration risks redundant inference and communication loops. Research directions include: (1) \textit{modality-specific energy budgets} that dynamically allocate computation to the most informative sensory channels; (2) \textit{shared semantic latent spaces} that enable efficient knowledge transfer across agents without full model replication; and (3) \textit{consensus-based distributed reasoning} that aggregates partial inferences rather than centralizing all computation. Hao \textit{et al.}~\cite{hao2022multiagent} demonstrate that multi-agent collaborative inference via DNN decoupling can reduce inference latency by up to 56\% and save up to 72\% of energy consumption, suggesting the potential for coordinated agent systems.

    \item 
\textbf{Reconciling green AI with performance requirements.} The pursuit of energy efficiency often conflicts with latency constraints and accuracy demands, particularly in safety-critical applications, e.g.,  autonomous driving or industrial robotics. Static trade-offs (e.g., fixed quantization levels) are insufficient; instead, context-aware performance contracts are needed that adapt the energy-accuracy-latency profiles based on task criticality. For instance, a navigation agent might employ full-precision reasoning at intersections but aggressive compression on open highways.
\end{itemize}

\subsection{Emerging Paradigms}

\begin{itemize}[leftmargin=*]
  \item \textbf{Federated green learning.} 
  Federated learning enables collaborative model improvement without raw data sharing, yet standard aggregation overlooks energy heterogeneity across devices, which is critical for multi‑agent systems coordinating reasoning and action at the edge. Lei \textit{et al.}~\cite{lei2024energy} propose sparse networks with an energy‑saving algorithm, cutting consumption by 56.21\% versus benchmarks and enabling sustainable collective intelligence. Chen \textit{et al.}~\cite{ni2025scheduling,chen2024toward} reduce energy through joint optimization of transmit power, bandwidth, compute, and scheduling, essential for maintaining the perception–reasoning–action loop without draining batteries. Future directions include: (1) \textit{energy‑aware federated optimization} weighting updates by battery and task criticality; (2) \textit{semantic gradient compression} transmitting only decision‑relevant updates under wireless energy budgets~\cite{wang2024survey}; and (3) \textit{asynchronous aggregation} accommodating intermittent agents while preserving coordination. Federated distillation (Section~\ref{sec:model-simp}) should further extend to multi-task agentic settings where agents share energy-efficient representations while learning specialized skills, enabling scalable autonomy without centralized energy costs.

  \item \textbf{Carbon-aware agency.} 
  Beyond device-level energy optimization, Agentic AI systems should incorporate operational carbon footprint as a first-class optimization objective. Recent advances in carbon-aware scheduling~\cite{chen2024joint,chen2025carbon,chadha2023greencourier} demonstrate how intelligent workload management can exploit temporal and geographic variations in renewable energy availability, potentially reducing emissions without changes to underlying models. Key requirements include: (1) \textit{carbon intensity forecasting} to schedule computation during periods of renewable energy abundance; (2) \textit{geographic load balancing} that routes inference requests to regions with green energy grids; and (3) \textit{carbon budgeting APIs} that allow applications to specify sustainability constraints alongside latency and accuracy requirements. Such carbon-aware scheduling transforms Agentic AI from an energy consumer to an active participant in grid decarbonization.
  
  \item \textbf{6G-enabled Agentic AI.} 
The convergence of 6G and Agentic AI enables native integration of communication, computation, and intelligence at the edge~\cite{zhou2024decentralized}. Unlike current overlay approaches where AI and networking operate in isolation, 6G enables semantic-aware network functions that understand and prioritize agentic workloads based on task semantics. Key directions include: (1) \textit{intelligent slicing for agentic workflows} that reserves bandwidth, compute, and energy by semantic priority~\cite{zhang2022workshop}; (2) \textit{AI‑native 6G protocols} embedding inference-aware semantics at the physical layer, where MCSs adapt to action criticality and transmission energy scales with the actionable value of information rather than raw bit count, preserving energy for high-stakes decisions while compressing routine perceptions. 
These advances transform 6G into an active participant in Agentic AI, enabling autonomous, collaborative, and energy‑efficient agents at scale.
  
  
  \item \textbf{Energy harvesting and self-sustaining systems.} 
  Batteryless IoT devices powered by energy harvesting (solar, RF, kinetic) offer a path toward perpetual operation~\cite{9069257}. Future Agentic AI systems must incorporate energy-adaptive inference that modulates computational depth based on available harvested energy, and task scheduling that aligns high-energy reasoning with periods of energy abundance. The ultimate vision is energy-positive agentic systems that not only sustain their own operation but contribute excess harvested energy back to the grid, transforming AI from an energy burden to a distributed energy asset \cite{chen2022distributed,chen2024SG}.
\end{itemize}


Looking forward, the ultimate goal is self-sustaining Agentic AI, referring to systems that continuously optimize their own energy efficiency through meta-learning and self-modeling. This involves: (1) \textit{energy-aware neural architecture search} that discovers models optimized for specific hardware-energy landscapes~\cite{bakhtiariifard2024ecnas}; (2) \textit{lifelong energy efficiency}, where agents learn to improve their inference efficiency over deployment time without forgetting; and (3) \textit{carbon-negative operation}, where intelligent workload scheduling and renewable energy harvesting combine to offset historical emissions. Achieving this vision requires cross-disciplinary collaboration spanning machine learning, wireless communications, hardware design, and sustainability science.

\section{Conclusion}
\label{sec:conclusion} 
This survey has systematically addressed the energy challenges of Agentic AI, whose closed-loop Perception-Reasoning-Action pipeline creates compounding computational and communication costs distinct from traditional AI systems. We have established a modular energy accounting framework revealing four distinct profiles (compute-bound Perception, memory-bound Reasoning, communication-bound Action, and bandwidth-bound Memory), and demonstrated that sustainable deployment requires coordinated optimization across model simplification, adaptive computation, input/attention efficiency, and hardware-aware inference. Cross-layer co-design emerges as essential, jointly tuning AI parameters with wireless transmission and edge resources to transform energy from an afterthought into a first-class constraint. Looking forward, federated green learning, carbon-aware agency, and 6G-native Agentic AI promise to embed sustainability into autonomous systems, enabling self-sustaining Agentic AI that dynamically adapts its resource usage to available energy budgets without compromising critical performance.

\bibliographystyle{ACM-Reference-Format}
\bibliography{references}

\end{document}